\documentclass[usenatbib,useAMS]{mn2e} 

\usepackage[T1]{fontenc}
\usepackage{ae,aecompl}
\usepackage{pdfpages}
\usepackage{epsfig,graphicx,float,xcolor}
\usepackage{subfigure}
\usepackage{multirow}
\usepackage{url}
\usepackage{color}
\usepackage{amsmath}
\usepackage{pdflscape}	
\usepackage{booktabs}

\usepackage{amssymb}
\usepackage{times}

\def\mnras{MNRAS}

\def\apj{ApJ}
\def\apjl{ApJL}
\def\apjs{ApJS}
\def\aj{AJ}
\def\jcap{JCAP}

\def\prd{PRD}
\def\aap{Astron. \& Astrophys.}

\def\physrep{Physics Reports}

\newcommand{\dd}{\mathrm{d}}

  \title[Cross-correlation of SDSS galaxies and clusters]{Cross-correlation of galaxies and
  galaxy clusters in the Sloan Digital Sky Survey  and the importance of non-Poissonian shot noise}

\author[Paech et al.]{Kerstin Paech$^{1,2}$\thanks{E-Mail: paech@physik.lmu.de},
  Nico Hamaus$^{1}$,
  Ben Hoyle$^{1,2,3}$,
  Matteo Costanzi$^{1}$,
  \newauthor
  Tommaso Giannantonio$^{1,4,5}$,
  Steffen Hagstotz$^{1}$,
  Georg Sauerwein$^{1}$,
  Jochen Weller$^{1,2,3}$ \\\\\\
$^1$Universit\"ats-Sternwarte M\"unchen, Fakult\"at f\"ur Physik, Ludwig-Maximilians-Universit\"at M\"unchen,
Scheinerstrasse 1, 81679 M\"unchen, Germany\\
$^2$Excellence Cluster Universe, Boltzmannstr. 2, D-85748 Garching, Germany\\
$^3$Max Planck Institute for Extraterrestrial Physics, Giessenbachstr. 1, D-85748 Garching, Germany. \\
$^4$Kavli Institute for Cosmology Cambridge, Institute of Astronomy, Madingley Road, Cambridge, CB3 0HA, United Kingdom\\
$^5$Centre for Theoretical Cosmology, DAMTP, University of Cambridge, Wilberforce Road, Cambridge CB3 0WA, United Kingdom
 }

\begin{document}
\label{firstpage}

\pagerange{\pageref{firstpage}--\pageref{lastpage}} \pubyear{2016}
\maketitle

\begin{abstract}
We present measurements of angular cross power spectra between galaxies and
optically-selected galaxy clusters
in the final photometric sample of the Sloan Digital Sky Survey (SDSS).
We measure the auto- and cross-correlations between galaxy and cluster samples,
from which we extract the effective biases
and study the shot noise properties. We model the non-Poissonian shot noise by
introducing an effective number density of tracers and fit for this quantity.
We find that we can only describe the cross-correlation of galaxies and
galaxy clusters, as well as the auto-correlation of galaxy clusters,
on the relevant scales using a non-Poissonian shot noise contribution.

The values of effective bias we finally measure for a
volume-limited sample are $b_{cc}=4.09 \pm 0.47$ for the cluster auto-correlation
and $b_{gc}=2.15 \pm 0.09$ for the galaxy-cluster cross-correlation. We find that
these results are consistent with expectations from the auto-correlations of galaxies
and clusters and are in good agreement with previous studies.
The main result is two-fold: firstly we provide a measurement of the cross-correlation
of galaxies and clusters, which can be used for further cosmological analysis,
and secondly we describe an effective treatment of the shot noise.

\end{abstract}

\section{Introduction}
%
The cosmological distributions of density and temperature perturbations are well approximated over
sufficiently large scales by Gaussian random fields, completely described by their two-point statistics.
One of the most powerful tools of modern cosmology is therefore the analysis of
{two-point} correlation functions, {which} can be measured as auto-correlations on one data set or as
cross-correlations between two data sets. The strongest current constraints on the cosmological model
are {indeed} derived from the measurement of the auto-correlation of the temperature
anisotropy of the cosmic microwave
background. Correlations can also be measured from the distribution of tracers of
the matter in the Universe: in the last decades multiple surveys have produced large galaxy catalogues,
 which allowed high-precision measurements of the
{galaxy} auto-correlation,  {such as
the two degree field (2dF) Galaxy Redshift Survey \citep{Cole:2005sx,Percival:2001hw}
   and the
SDSS \citep{York2000,Tegmark:2003ud,Hayes:2011aa,Ho:2012aa,Beutler:2014aa,Grieb:2016aa}}.
{Likewise,} the availability of large optically-selected galaxy cluster catalogues has led to the
measurement of the auto-correlation of galaxy clusters, e.g.
from the SDSS catalogue \citep{Huetsi:2009zq,Estrada:2008em,Miyatake:2016aa,Baxter:2016aa,Veropalumbo:2016aa},
{and from the REFLEX X-ray survey}
\citep{Collins:2000yk,Balaguera-Antolinez:2011aa}. These measurements have also been used
to obtain cosmological constraints, for both
{the REFLEX catalogue} \citep{Schuecker:2003aa}
{and} several cluster samples from the SDSS, such as maxBCG \citep{Mana:2013qba}.

Given the success of auto-correlation measurements and the abundance of
different cosmological probes of the density field, it is increasingly
interesting to combine probes
via cross-correlations. {Cross-correlations, such as
  for example between galaxy surveys and the cosmic microwave
background (CMB) temperature and lensing \citep{Giannantonio2014,Giannantonio:2016aa},
or between galaxies and cosmic voids \citep{Hamaus:2014aa,Hamaus:2016aa},
provide new information without
requiring new observations, and can thus lead to improved and complementary cosmological constraints. }

Some measurements of cross-correlation between galaxy clusters and galaxies
were attempted
in the 1970s and 1980s \citep{Peebles1974ApJS,SeldnerPeebles1977Apjla,
SeldnerPeebles1977ApJb,LiljeEfstathiou1988MNRAS}. These studies were performed
on relatively small and non-independent catalogues:
the cluster catalogues used by all groups were drawn from \citet{Abell1958}
and the galaxy catalogues were either the galaxy
counts by \citet{Shane1967} or by \citet{1977AJ.....82..249S}.
The better of these two galaxy catalogues had a
resolution of 10 arcmin $\times$ 10 arcmin on about $19\deg^2$. These
early cross-correlation analyses were therefore limited in their possible applications.
Some more recent works measuring galaxy and galaxy-cluster cross-correlations
are \citet{Croft:1999aa,2005MNRAS.362.1225S,2013MNRAS.431.3319Z}.

\citet{Hutsi:2008aa}
proposed the measurement of the correlation between
galaxy clusters and galaxies as an additional cosmological probe, which
was later extended by \citet{2011MNRAS.414.1545F}.
They showed that the cross-correlation of clusters and galaxies
could lead to better constraints on cosmological parameters,
{as well as a better determination of} the halo model parameters \citep{Cooray:2002aa}.

In this paper, we measure the cross-correlation between galaxies and
clusters derived from the final
photometric data release of SDSS (Data Release 8, DR8)  \citep{Aihara:2011sj}.
When  using linear theory and cluster bias, as well as Poissonian shot noise, we
find a discrepancy between the theoretical expectations and the measured
angular power spectra. We show that this tension can be resolved by adopting a
modified treatment of the shot noise.

The outline of this paper is as follows:
we describe in Section~\ref{sect:theory} the theoretical modelling of the angular
power spectra, the shot noise, and the cluster bias.
In Section~\ref{sect:data} we introduce the catalogues and mask used in
the analysis, and in Section~\ref{sect:measuring} we present the details of
the angular power spectra $C_l$ estimation.  Section~\ref{sect:analysis} presents
the results for the auto- and cross-correlations of galaxies and galaxy clusters.
Finally,  our summary and outlook are given
in Section~\ref{sect:conclusions}.

\section{Theoretical Modelling of the Angular Power Spectra}
\label{sect:theory}

In order to extract cosmological parameters from the measured galaxy and
cluster angular power spectra $C^{\rm data}_l$, we need theoretical model
predictions $C^{\rm model}_l$ that account for systematics and measurement
effects affecting the observed correlation functions.

\subsection{Angular power spectra of biased tracers}
\label{class_theory}
We define the density field of {the mass density fluctuations} at
comoving coordinate $\mathbf{r}$ {and at any redshift $z$} as
\begin{equation}
\delta_{m}(\mathbf{r}) = \frac{\rho_{m}(\mathbf{r})}{\bar{\rho}_{m}} - 1\;,
\end{equation}
where $\rho_{m}(\mathbf{r})$ is the spatially varying matter density in the Universe
with a mean of ${\bar \rho_{m}}$. {The matter overdensity $\delta_{m}$} can be related to the galaxy ({or} cluster)
{overdensity $\delta_{a}$ (where $a$ denotes a galaxy or cluster sample)} via the local bias model \citep{Fry:1993aa},
\begin{equation}
\delta_{a}(\mathbf{r}) \simeq b_{1,a}\delta_{m}(\mathbf{r}) +
\frac{b_{2,a}}{2}\delta_{m}^2(\mathbf{r}) + O(\delta_{m}^3) + \varepsilon_a \;,
\end{equation}
{with linear and non-linear bias parameters $b_{1,a}$, $b_{2,a}$, and a shot noise term $\varepsilon_a$}.

{In Fourier space, we can define the matter, galaxy, or cluster power spectra between any pair of samples $(a,b)$ as:}
\begin{equation}
(2\pi)^3 \delta_{\rm D}(k-k^\prime) P_{ab} (k) \equiv \langle \delta_{a}(\mathbf{k}) \, \delta^{\star}_{b}(\mathbf{k^\prime}) \rangle\;,
\end{equation}
where $\mathbf{k}$ denotes a wave vector of amplitude $k$ and angled brackets indicate an
average over all Fourier modes within a given spherical shell and $\delta_{\rm D}$ is
the Dirac delta function. Up to linear order and assuming
Poissonian shot noise, the galaxy (or cluster) power spectrum can be directly related to the
matter power spectrum $P(k)$
,
\begin{equation}
\label{eq:shot}
P_{ab}(k) \simeq b_{1,a}b_{1,b} \, P(k) + \delta^{ab}_{K} \, V/N_a
\end{equation}
where $\delta_K$ is the Kronecker delta, and the shot noise contribution is given
by the inverse number density of galaxies (or clusters), $V/N_a$.

In this analysis we consider the angular power spectrum $C^{ab}_l$, a projection of $P_{ab}(k)$
on the sky.
We use the publically available code CLASS\footnote{\mbox{http://class-code.net/}}~\citep{Blas:2011aa}
to generate theoretical predictions for
the angular cluster power spectrum. CLASS is a differential equation solver for the hierarchy
of  Boltzmann equations governing the perturbations in the density of dark matter,
baryons, photons and any other relevant particle species. The CLASSgal
extension \citep{Di-Dio:2013aa} calculates the angular power spectrum, $C^{ab}_l$, for any matter tracer as
\begin{equation}
\label{eq:class_cl}
C^{ab}_l = 4 \pi \int \frac{\rm d k}{k} P_{\rm ini}(k) \Delta_l^a(k)\Delta_l^b(k) \: ,
\end{equation}
where $P_\mathrm{ini}$ denotes the (dimensionless) primordial power spectrum and the transfer
function for the matter component $\Delta^a_l(k)$
is given by
\begin{equation}
\Delta^a_l(k) = \int {\rm d} z \: b_{1,a} \frac{ {\rm d} N_a}{{\rm d} z} j_l(k r(z)) {\mathcal{D}}(k,z) \: .
\end{equation}
Here $r(z)$ is the comoving distance, ${\mathcal{D}}(k,z)$ the total comoving density
fluctuation\footnote{${\mathcal{D}}(k,z) \approx D_+(z)T(k)$ for cold dark matter universes, where
$D_+(z)$ is the density growth function and $T(k)$ is the matter transfer function.}
and we use the galaxy and
cluster redshift distributions ${\rm d} N/{\rm d} z$ for the observed sample,
which are shown in Figure~\ref{zdist} and introduced in Section~\ref{sect:data}.

The main goal of this analysis is to measure the auto- and cross-correlation
of galaxies and clusters, and to determine the effective bias of
these tracers.
Therefore we fix the cosmological parameters to their best-fit values
{as obtained by the \textit{Planck} collaboration \citep{Planck-Collaboration:2014aa}
(Planck2013+WP+highL+BAO),
derived by combining their own CMB data with
the Wilkinson Microwave Anisotropy Probe (WMAP) polarization data
\citep{Bennett:2013aa},
the small-scale CMB measurements from the Atacama Cosmology
Telescope (ACT) \citep{Das:2014aa} and the South Pole Telescope (SPT)
\citep{Reichardt:2012aa}, as well as
baryonic accoustic oscillations (BAO) data from SDSS}
\citep{Percival:2010aa,Padmanabhan:2012aa,Blake:2011aa,Anderson:2012aa,Beutler:2011aa}.
The cosmological parameters we use are: $h=0.678$, $\Omega_{\rm b}=0.048$
$\Omega_{\rm c}=0.258$, $\sigma_8=0.826$, $z_{\rm re}=11.3$ and $n_{\rm s}=0.96$
(we checked that assuming a Planck 2015 cosmology has no significant impact on the
results in our analysis).

For the {analysis presented in this paper, we adopt a constant bias model,
i.e. we define an effective bias $b_{\rm {eff}}$, such that we can assume for
each sample $b_{1,a}(z) = b_{\rm{eff,a}}$.
The full redshift evolution of the galaxy and cluster bias could in principle
be obtained by subdividing our samples in multiple redshift bins, but this is
beyond the scope of the present analysis and the data available.}


Note that the CLASS $C^{ab}_l$ do not account for a contribution due shot noise.
As we demonstrate below, the theoretical power spectra $C^{ab}_l$ defined
by Equation~(\ref{eq:class_cl})
need a more advanced modelling of the shot noise contribution, which we
present in the next Section.

\subsection{Accounting for shot noise}
\label{sect:shot_noise}
Estimating the underlying, continuous dark matter density field via the
{discrete} number density of observed galaxies and clusters, introduces a shot noise contribution
which will leave a systematic imprint on the measured angular power spectrum $C^{\rm data}_l$.
In real space, the Poisson sampling
from the true underlying density distribution introduces a contribution to
the auto-correlation at zero separation, which translates into the constant
contribution in harmonic space shown in Equation~(\ref{eq:shot}).
Due to the large number of galaxies observed in the SDSS DR8, this contribution
to the measured $C_l^{\rm data}$ is negligible on the relevant scales
for galaxies,
but is the leading contribution for the cluster auto-correlation function.

The situation is
more complicated for the galaxy-cluster cross-correlation.
Galaxies that are part of a cluster contribute to the  shot noise,
while those that are not part of a cluster do not.
Since the majority of the galaxies in our sample are not part of a galaxy cluster,
we set the shot noise contribution for the galaxy-cluster cross-correlation to zero
for now, but we will revisit this {issue} in Section~\ref{sect:fitting_n}.

Additionally, we have to consider a similar, although smaller, effect
for the cross-correlation of clusters in different richness bins.
Assuming these clusters occupy halos of different mass, self-pairs are not
taken into account in their cross-correlation,
resulting in a vanishing Poisson shot noise contribution.
We will come back to this issue in Section~\ref{sect:fitting_n} as well.

%

The Poisson noise contribution to the model power spectra $ C_l^{ab, \rm model}$
can be approximated by
\begin{equation}
{N}_l^{ab} =  \delta_K^{ab} \, f_{\rm sky}\frac{4\pi}{N_a} \quad ,
 \label{eq:shot_noise_approx}
\end{equation}
where $f_{\rm sky}$ is the fraction of the sky covered and $N_a$ is the
number of objects observed.

While Equation~(\ref{eq:shot_noise_approx}) holds for regular masks, in the
case of irregular
masks (as the one used in this analysis) a more accurate estimation of
the shot noise component is required.
In this case, in all generality the
shot noise contribution ${\widetilde N}_l$  can be determined by
Poisson sampling different random realisations of a sky map with a
constant matter density.

Each random realisation $i$ has a power spectrum $C^{{\rm rand},i}_l$,
from which an estimate of the shot noise contribution can be obtained by averaging:

\begin{equation}
{\widetilde N}_l = \langle C^{{\rm rand},i}_l \rangle \quad ,
 \label{eq:shot_noise_sampled}
\end{equation}
where the angular bracket $\langle \cdot \rangle$ denotes the average over all random
maps $i$. The covariance between different angular wave numbers $l$ and $m$ is given by
\begin{equation}
{\rm Cov}[C^{{\rm rand}}_l,C^{{\rm rand}}_m] =
\frac{N_{\rm s}}{N_{\rm s}-1}
\langle (C^{{\rm rand},i}_l - {\widetilde N}_l)
(C^{{\rm rand},i}_m - {\widetilde N}_m)\rangle
\label{cov_cl_poisson_random}
\end{equation}
where $N_{\rm s}=100$ is the number of samples used.
From this we can determine the covariance of the shot noise, ${\widetilde N}$,
as
\begin{equation}
{\rm Cov}[{\widetilde N}_l,{\widetilde N}_m] = N^{-1/2}_s \, { \rm
  Cov}[C^{{\rm rand},i}_l,C^{{\rm rand},i}_m] \quad .
\label{eq:neff_cov}
\end{equation}
We discuss how  ${\rm Cov}[{\widetilde N}_l,{\widetilde N}_m]$ enters
the analysis in Section~\ref{sect:fitting_nfixed}.


For full sky coverage, we recover the shot noise contribution
as given by Equation~(\ref{eq:shot_noise_approx}), which remains a good
  approximation as long as
  the shape of the mask is regular enough;
 we expect however to observe signifiant deviations for increasingly irregular masks.

The amplitude of the shot noise contribution ${\widetilde N}_l$ depends
on the number of objects $N_a$ distributed over the area of the mask. However,
the shape of ${\widetilde N}_l$ for different $l$ is independent of $N_a$, i.e.
for a given mask and pixel size, we can determine the shot noise
contribution just once then rescale the result according to the actual
number of objects observed.

Since we are working with pixellated maps as described in Section~\ref{sec:mask}
we will be using the average object per pixel density $\bar n$ when
determining the shot noise contribution.
The shot noise contribution then is determined as
\begin{equation}
{\widetilde N}_l (\bar n ) = {\widetilde N}_l
(1)/\bar n  \quad .
\label{eq:noise_const}
\end{equation}
This means that, for a given mask and pixel size, the shot noise contribution
can be determined once for a fixed
object per pixel density $\bar n = 1$ and then
rescale the result according to the actual object per pixel density
of our sample $\bar n$.
We discuss the actual shot noise contribution for the sky mask used in this
analysis in Section~\ref{sec:mask}.

\subsection{Sub- and super-Poissonian shot noise}
\label{sec:non_poissonian_shot_noise}
We expect deviations from a purely Poissonian shot noise contribution for the power
spectra when measured on the galaxy cluster data.
$N$-body simulations have provided significant evidence for {such} deviations
in the clustering statistics of dark matter
  halos~\citep{Hamaus:2010aa}.  In particular, these deviations have been shown
  to depend on halo mass: on large scales the shot noise contribution to
  the power spectrum of low-mass halos exceeds the fiducial value of
  $V/N_{a}$,
  while it is suppressed compared to that value at high masses. {These effects are}
  commonly referred to as \emph{sub-} and \emph{super}-Poissonian shot noise,
  respectively.  In addition, the Poisson expectation is not only found to be
  violated in auto-correlations of a single tracer, but also in
  cross-correlations among different tracers. While Poissonian shot noise only
  affects the auto-correlation of self-pairs, simulations have revealed
  non-vanishing shot noise contributions in cross-correlations between halos of
  different mass~\citep{Hamaus:2010aa}. These can be either positive or
  negative, depending on the considered mass ranges.

This phenomenology can be explained with two competing effects:
exclusion and non-linear clustering~\citep{Baldauf:2013aa}. The former simply
specifies the fact that any two tracers, be it halos or galaxies, can never be
closer to one another than the sum of their own extents. This violates the
Poisson assumption, which states that tracers are randomly sampled at any given
point within some volume. Each tracer contributes an exclusion region that
effectively diminishes the available sampling volume, and therefore the shot
noise contribution. As the exclusion region of halos increases with their mass,
high-mass halos are most affected by this. Moreover, the exclusion mechanism
also applies for halos of different mass, i.e.\ it influences cross-correlations
of tracers as well.

The first-order contribution from non-linear clustering of halos beyond linear
theory is described by the second-order bias parameter $b_2$.
Besides modifying
the scale-dependent linear clustering power spectrum of halos on small scales, it
also contributes a scale-independent term that cannot be distinguished from
Poisson shot noise~\citep{McDonald:2006aa}. Hence, non-linear
clustering effectively increases the Poisson shot noise, and this effect is most
important for low-mass halos, where the value of $b_2$ is non-zero and exclusion
effects are small.

While the above effects mainly apply to dark matter halos, they can be
translated to galaxies and clusters by means of the halo model
\citep{Seljak:2000aa,Smith:2003aa}. Given a Halo
Occupation Distribution (HOD), one can assign central and satellite galaxies to
each halo of a given mass. While centrals and cluster centers closely obey the
effects outlined above, satellites add more complexity as they do not
obey halo exclusion. In this case the satellite fraction determines the shot
noise as well: a low value ($\sim 5\%$) results in sub-Poissonian, and a high
value ($\sim 8.5\%$) in super-Poissonian shot noise~\citep{Baldauf:2013aa}.

\subsection{Accounting for sub- and super-Poissonian shot noise}
\label{sect:effective_n}
We should therefore expect deviations from the Poisson shot noise predictions on
all scales, and we expect this correction to be most important
for the cluster auto- and cluster-galaxy cross-correlations.

As we have discussed in the previous section, on
large scales we expect  the shot noise correction to be independent
of $l$; we can then model this by introducing
an inverse effective (average)
  number density $\bar n_{\rm eff}^{-1}$ free parameter, which we will fit from the data.
In this case, we replace $\bar n$
in Equation~(\ref{eq:noise_const}) with $\bar n_{\rm eff}$ and use the following
relation for the effective shot noise:
\begin{equation}
 {\widetilde N}^{\rm eff}_l (\bar n_{\rm eff} ) = {\widetilde N}_l
(1) \cdot \bar n_{\rm eff}^{-1}  \quad .
\label{eq:noise_eff}
\end{equation}
in the case of Poissonian shot noise we should recover $\bar n_{\rm eff} \simeq \bar n$.
We determine $n_{\rm eff}$ for
each of the auto- and cross-correlations and marginalise over it to determine
the effective bias {of each sample.} 

\subsection{Theory predictions for effective bias}
\label{sect:theory_bias}
We compare the effective bias we extract from the data to theoretical
expectations for a volume-limited sample, whose details are described below in Section~\ref{sect:vlim}.
We assume here the halo mass function $n(M,z)$ and halo bias $b(M,z)$ to be given by the fits
to $N$-body simulations by \citet{Tinker:2008aa, Tinker:2010aa}.
In order to calculate
the expected effective bias of a volume-limited cluster sample, we average over
the redshift range considered
\begin{eqnarray}
  b_{\rm eff} &=& \left[ \int_{\Delta z} \dd z \frac{\dd V}{\dd z\dd\Omega} \int \dd M n(M,z) b(M,z) \right.  \\
&&  \hspace{3cm}\left.  \int_{\lambda_{\rm min}} \dd\lambda P(\lambda |M,z)\right] \nonumber \\
&& \bigg/ \left[ \int_{\Delta z} \dd z \frac{\dd V}{\dd z \dd\Omega}  \int \dd M n(M,z)
    \int_{\lambda_{\rm min}} \dd \lambda P(\lambda |M,z) \right] \, , \nonumber
\end{eqnarray}
where $\dd V/(\dd z\dd\Omega)$ is the comoving volume element per unit redshift
and solid angle,
and $P(\lambda |M,z)$ is the
probability that a halo of mass $M$ at redshift $z$ is observed with a richness
$\lambda$. We model this probability with a log-normal distribution:
\begin{equation}
 P(\lambda | M,z) = \frac{1}{\lambda \sqrt{2 \pi \sigma_{\ln \lambda}^2}} \exp
 \left[ -\frac{\ln \lambda - \langle \ln \lambda(M,z) \rangle}{2 \sigma_{\ln
       \lambda}^2} \right] \, ,
\end{equation}
where we use the parameterisation and parameter values of the scaling
relation between mass and richness $ \langle \ln \lambda(M,z) \rangle$
as determined by~\citet{Farahi:2016aa}.
The scatter $ \sigma_{\ln \lambda}$ is defined as
\begin{equation}
  \sigma_{\ln \lambda}^2 =\frac{\exp \left[ \langle \ln \lambda(M,z) \rangle \right] - 1}
        {\left(\exp \left[ \langle \ln \lambda(M,z) \rangle \right] \right)^2}
        + \sigma_{\ln\lambda|M}^2 \, ,
\end{equation}
where the first term accounts for the richness-dependent Poisson noise and
$\sigma_{\ln\lambda|M}$ is the intrinsic scatter in the richness-mass relation.
As~\citet{Farahi:2016aa} do not specify the value for
$\sigma_{\ln\lambda|M}$ or $\sigma_{\ln M|\lambda}$, we determine
$\sigma_{\ln\lambda|M}$ using the value for $\sigma_{\ln M|\lambda}=0.25$
from~\citet{Simet:2016aa}.
Using Equations (13) to (15) from \citet{Simet:2016aa},
we determine  $\sigma_{\ln\lambda|M}$
via the following relation
\begin{equation}
 \sigma_{\ln\lambda|M} = \frac{ \sigma_{\ln M|\lambda} }{\alpha} \quad,
\end{equation}
where $\alpha=1.326$ denotes the power-law slope of mass given the richness,
as defined by~\citet{Simet:2016aa}.

\section{Data}
\label{sect:data}

\subsection{Galaxy and cluster catalogues}
\label{sec:cats}
We use galaxy and galaxy cluster data drawn from the Sloan
Digital Sky Survey (SDSS) \citep{York2000}. The SDSS is conducted with a dedicated 2.5m telescope
at the Apache Point Observatory in Southern New Mexico in the United
States. This telescope has a wide field of view of $7 \deg^2$, a large mosaic
CCD camera and a pair of double spectrographs \citep{Aihara:2011sj,
  Eisenstein:2011sa}. We use the final photometric SDSS data from the eighth data release (DR8) that
combines data from the two project phases SDSS-I and SDSS-II.

The full area of SDSS DR8 is $14,555 \deg^2$ and includes photometric
measurements of 208,478,448 galaxies. For the analysis in this paper we use
the same galaxy catalogue and selection
criteria as in \citet{Giannantonio:2012aa}. The catalogue only contains objects
with redshift between 0.1 and 0.9 that have a photo-$z$ uncertainty of $\sigma_z(z) < 0.5 \, z$.
A completeness cut is applied by only using objects with extinction-corrected $r$-band magnitudes between 18 and 21.  After these cuts, the catalogue
contains 41,853,880 galaxies.

We use the cluster catalogue
 constructed from SDSS DR8 with the
\textbf{red}-sequence \textbf{Ma}tched-filter \textbf{P}robabilistic
\textbf{Per}colation (redMaPPer) cluster finding
algorithm version 5.10\footnote{  \mbox{http://risa.stanford.edu/redmapper/}} \citep{Rykoff:2013ovv}.
It contains $\sim 26,350$ galaxy clusters covering a redshift range between 0.1 and 0.6
and contains only clusters with richness $\lambda>20$. Note that we use the richness
as defined by redMaPPer throughout this paper. As the richness
$\lambda$ is a measure of the number of galaxies within the cluster, this means
smaller clusters, which are more strongly affected by systematic errors, are excluded
from the analysis.

\subsection{Pixellated maps and survey mask}
\label{sec:mask}
We create pixelized maps for the galaxy and the galaxy
cluster catalogues using the pixelization scheme
HEALPix\footnote{\mbox{http://healpix.jpl.nasa.gov/}}
\citep{Gorski:2004by}, in which the resolution is expressed
by the parameter $N_{\rm side}$.
The pixellation effectively
smoothes information on scales
smaller than the pixel size, and it can be described by a multiplicative
window function $w_l$ given by the pixel window function provided by
HEALPix.

To construct the (binary) survey mask we follow the method by
\cite{Giannantonio:2006aa} to estimate the coverage of pixels that straddle the survey boundaries.
If we have a distribution $P(n)$ of the number of galaxies per
pixel, this effect causes a deviation from a Poisson distribution  $P^{\rm Poiss}(n)$
for low $n$, i.e.\ we will find an excess of pixels
with a small  $n$  compared to $P^{\rm Poiss}(n)$.
In practice, we create the mask by discarding pixels where
$P(n) \not\approx P^{\rm Poiss}(n)$, i.e.\ pixels with $n < n_{\rm min}$
where $n_{\rm min}$ is a cut-off threshold we choose.

The pixel size we choose is constrained by two factors.
On the one hand pixels
should be large enough to ensure that the mean of the distribution $P(n)$
is far from zero, i.e. there is only a very small and negligible number of pixels
with a small number of galaxies.
On the other hand pixels should
be small enough so that we can  measure the angular power spectra at the
scales of interest.

We choose the HEALPix resolution $N_{\rm side} = 512$, corresponding to a
pixel side of $7'$, as this produces an average number of galaxies per
pixel of $\bar n \simeq 30$ for the
SDSS galaxy catalogue, and it also allows access to the scales of interest
in this analysis. All catalogues we use here are pixelized at this resolution.

Therefore, to create the mask we first
determine the number of galaxies $n_i$ in each of the pixels $i$ and discard
pixels with zero galaxies.
Then we determine the mean and variance of the Poisson distribution
by calculating the  average number of galaxies per pixel $\bar n$ and
identify the value of $n$ below which we observe a deviation from the Poisson distribution.
We find that $n_{\rm min}  = \bar n - 2 \sqrt{\bar n}$ (i.e.\ the equivalent
of two standard deviations below the mean number of pixels) is a good value
for the cut-off and we mask all pixels where $n_i < n_{\rm min}$.

We generate the cluster mask by repeating the above process using
the available random redMaPPer cluster catalogues.
We choose to only work with the intersection of
the galaxy and galaxy cluster masks.
We further combine this mask with the dust extinction maps
by~\citet{Schlegel:1998lr}, retaining only pixels with reddening values $E(B-V) < 0.2$,
and with the SDSS seeing masks by
\citet{Ross:2011aa},
retaining
pixels with seeing values below $1.4''$. The final mask footprint covers
6,983~deg$^2$.
After the application of the complete mask, the data covers
a fraction of sky $f_{\rm sky}=0.21$ and contains 671,533 unmasked pixels.

\begin{figure}
  \hspace*{-1cm}
	 \centering \includegraphics[scale=0.45]{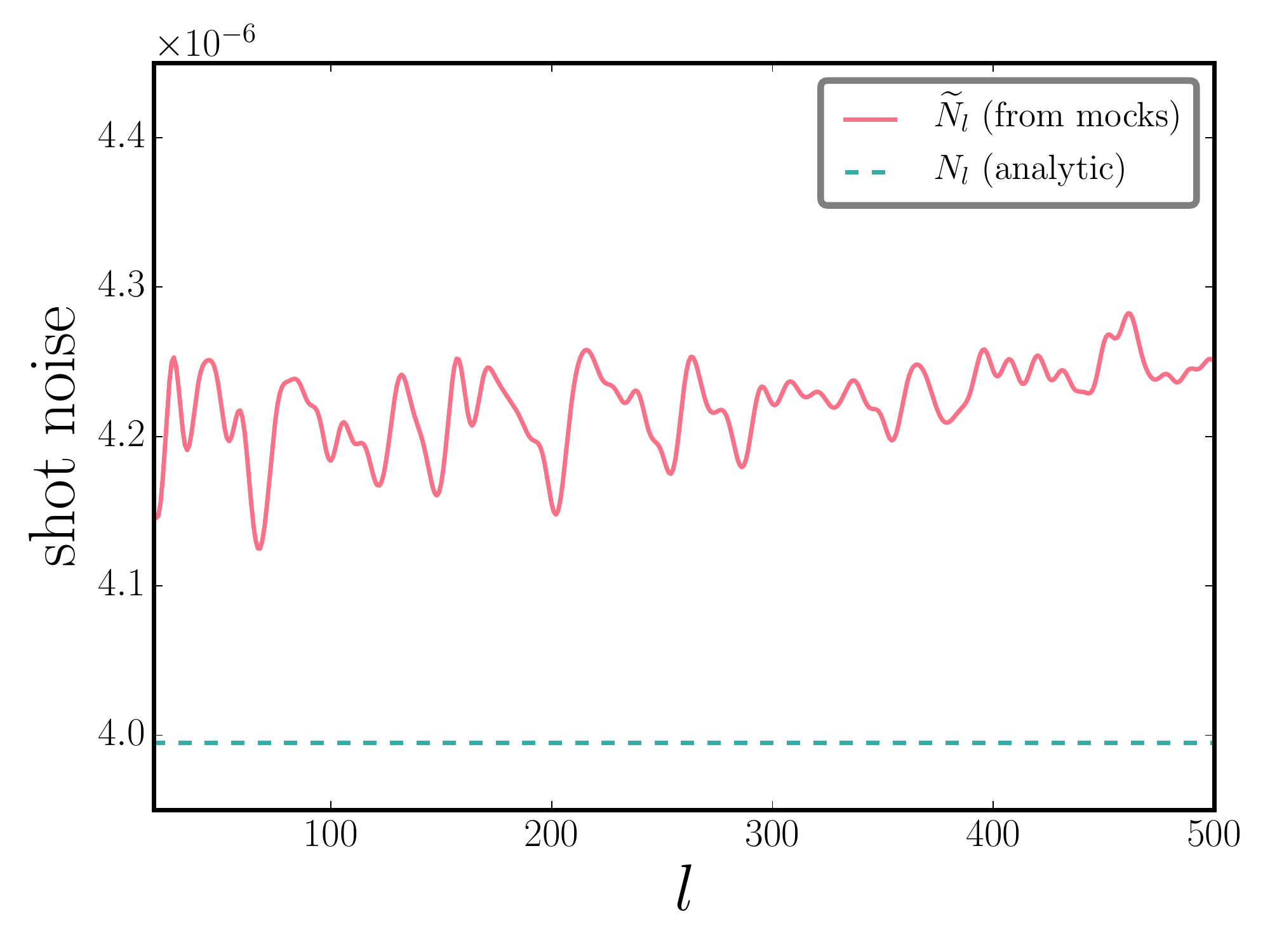}
	\caption{ Shot noise predictions for an
          approximate {\em analytic} shot noise estimate ${N}_l$
          (Equation~\ref{eq:shot_noise_approx}) and the Poisson sampled shot noise
          ({\em mocks})
          ${\widetilde N}_l$ (Equation~\ref{eq:shot_noise_sampled}) for an average
            number of objects per pixel density $\bar n = 1$. }
          \label{fig:noise_cl}
\end{figure}

In Section~\ref{sect:shot_noise} we described how we determine the shot noise
contribution for a given mask.
Figure~\ref{fig:noise_cl} shows both the analytic shot noise
approximation ${N}_l$ from Equation~(\ref{eq:shot_noise_approx})
and the shot noise ${\widetilde N}_l$ from Poisson sampling for $\bar n = 1$
from Equation~(\ref{eq:shot_noise_sampled}) for the mask used in our analysis.
We see that taking
into account the effect of the mask increases the shot noise by approximately~5\%.
Also, there is a mild $l$-dependence of the shot noise for the Poisson
sampled shot noise compared to the analytic approximate shot noise
caused by the shape of the mask.

\subsection{Subsets of data used}
\label{sect:data_samples}
In addition to the full sample described in Section~\ref{sec:cats} above,
for our analysis we use different subsets of the cluster catalogue. In the following, we will consider:
\begin{itemize}
\item $c_{\rm all}$: the full cluster sample, which is richness-selected and thus not volume-limited;
\item $c_{\rm vlim}$: a volume-limited sample that is constructed by using only clusters with $z<0.35$;
\item $c_{\lambda_{\rm low}}$ and  $c_{\lambda_{\rm high}}$: a low- and high-richness sample,
  constructed from the full cluster sample (all redshifts) that is split at the
  median richness of $\lambda_{\rm med} = 33.7$.
\end{itemize}
The different samples are summarised in Table~\ref{tab:samples}.
The sample $c_{\rm all}$ containing all the clusters of the redMaPPer catalogue
is the starting point of our analysis as it makes use of all
the objects available and allows us to investigate the shot noise properties of our
measurements, especially for the galaxy and galaxy cluster cross-correlation
(Sections~\ref{sect:fitting_nfixed} and~\ref{sect:fitting_n}).
In a second step we  use the $c_{\rm vlim}$ sample
in order to be able to compare our best-fit results to the theoretical
expectation for the value of the effective bias (Section~\ref{sect:vlim}).
Finally, we use the samples split into two richness bins
$c_{\lambda_{\rm low}}$ and  $c_{\lambda_{\rm high}}$ to investigate the shot noise properties of
cluster clustering (Section~\ref{sect:richness_fits}).

\begin{figure}
\hspace*{-1cm}
	\centering \includegraphics[scale=0.45,  
          clip=true]{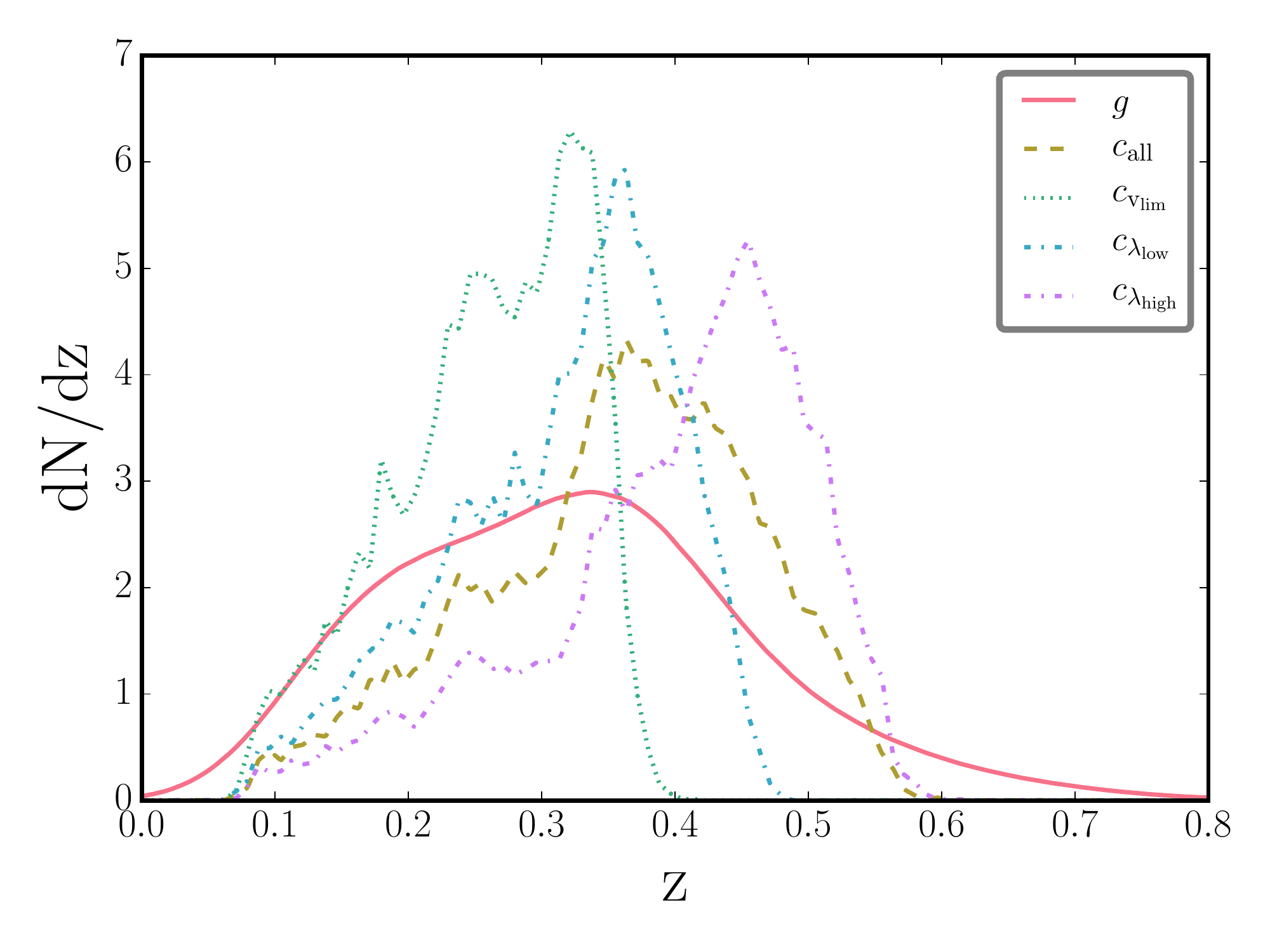}
	\caption{Normalized redshift distributions ${\rm d}N/{\rm d}z$
        of galaxies
          and galaxy clusters samples as
          discussed in Section~\ref{sect:data_samples}.}
        \label{zdist}
\end{figure}

\begin{table}
  \begin{center}
  \caption{
    Summary of all samples used in this analysis. $N$ is the total number of
    objects left after masking and cuts, $\bar n$ is the average number of
    objects per pixel ($N_{\rm side}=512$), $z_{\rm median}$ is the
    median redshift of the sample and the selection column indicates the
    additional cuts done beyond masking.
  }
  \label{tab:samples}
  \begin{tabular}{cccccc}
    \toprule
    Sample & $N$  & object type  & $z_{\rm median}$ & $\bar n$ & selection \\
    \midrule
    g & 25,959,346 & galaxies & 0.31 & 39 & see text\\
    $c_{\rm all}$ & 21,962 & clusters & 0.37 & 0.033 & none \\
    $c_{\rm vlim}$ & 9,294 & clusters & 0.27 & 0.014 & $z<0.35$ \\
    $c_{\lambda{\rm low}}$ & 10,981 & clusters & 0.34 & 0.016 & $\lambda<33.7$ \\
    $c_{\lambda{\rm high}}$ & 10,981 & clusters & 0.42 & 0.016 & $\lambda>33.7$ \\
    \bottomrule
\end{tabular}
\end{center}
\end{table}

In Figure~\ref{zdist} we show the redshift distribution ${\rm d}N/{\rm d}z$ of galaxies and galaxy clusters,
after masking has been applied,
for the different samples described above.
In order to account for the uncertainties in cluster redshifts when performing the
theoretical predictions in Sections~\ref{class_theory}, we randomly sample from the
redshifts errors for the respective object type. For galaxies we assume an overall
5\% photometric redshift error as the individual redshift errors in the
catalogues underestimate the true redshift error. In the case of the galaxy
clusters we re-sample the provided redshift according to the error provided
by redMaPPer.
Therefore the redshift
distribution of galaxies does not have a sharp boundary at low redshifts
and the distribution of clusters for the $c_{\rm vlim}$ sample does not have
a sharp boundary at
$z = 0.35$.

\section{Angular power spectrum estimators}
\label{sect:measuring}

We measure the angular power spectrum $C_l$ for
all the data products described in Section~\ref{sect:data} using the
Spatially Inhomogeneous Correlation Estimator for Temperature and Polarisation
(PolSpice)\footnote{\mbox{http://www2.iap.fr/users/hivon/software/PolSpice/}}
\citep{Chon:2003gx, Szapudi:2000xj,Challinor:2005aa}.
The advantage of using PolSpice is that the algorithm corrects for distortions
of the measured power spectrum caused by masking and the pixel window
function $w_l$.
The partial sky coverage has more complex effects \citep[see eg][]{Efstathiou:2004aa},
which we assume here to be corrected by PolSpice.

PolSpice is used to estimate the $C^{\rm data}_l$ from pixellated density contrast maps
$\delta_i$ derived from both the galaxy and galaxy cluster density per pixel~$n_i$
\begin{equation}
\delta_i = \frac{n_i}{\bar n} - 1 \, ,
\end{equation}
where $\bar n $ denotes the average over all pixels $i$.

For our analysis we consider multipoles $l$ in the range $20 < l < 500$.
The minimum multipole is limited by the size of the mask and we choose a
cautious estimate following~\citet{La-Porta:2008aa}. For the maximum
multipole we choose an equally conservative limit by using $l < N_{\rm side}$
instead of $l<2N_{\rm side}$ as reasoned by \citet{Gorski:2004by}.

Following the argument by \citet{2012MNRAS.422...44P} (and
references therein and their Figure 7) we expect the onset of the bias non-linearity
at $k > 0.2\,h$/Mpc, which
corresponds to a multipole $l\approx200$ at a mean redshift similar to
the mean redshift of the cluster samples described
in Section~\ref{sect:data_samples}.
However, while limiting our analysis to $l<200$ does not change the results
significantly, the statistical power of our results is reduced noticeably.
Therefore we will use $l < N_{\rm side}$ throughout this paper and
perform the cross-check described in Section~\ref{sect:fitting_nfixed}
to ensure that the non-linearities do not have a notable effect on our results.

To some extent the non-linearities can be absorbed by the way we treat
the shot noise, however this cannot account for the entire mass, scale and
redshift dependence of the non-linear bias.
Nevertheless, the cross-check discussed in
Section~\ref{sect:fitting_nfixed} gives us confidence that
any residual non-linearities are sub-dominant, because the scale-dependent bias of
tracers with different host halo mass cannot be accounted for in our model
and would lead to tensions in the cross-check.

We estimate the covariance of the $C_l^{\rm data}$ by using a jack-knife
sampling of the maps, which is performed by dividing the map
area into $N_{\rm jk}$ regions of
equal size. For each sampling $i$ one of the regions is left out and the
measurement of the $C^i_l$ is done on the area constituted by the remaining
$N_{\rm jk}-1$ regions.
The covariance matrix using jackknife sampling is then given by
\begin{equation}
{\rm Cov}^{\rm data}[C_l,C_m] =  \frac{N_{\rm jk} - 1}{N_{\rm
    jk}} \sum\limits_{j}{(C^i_l - \langle C_l \rangle)(C^i_m - \langle C_m
  \rangle)} \, .
\label{cov_data}
\end{equation}

The number of jack-knife samples $N_{\rm jk}$ needs to be large enough
to determine the covariance matrix with sufficient accuracy for a given number of
data bins $N_{\rm  bin}$ in $l$-space. Following the reasoning by \cite{Taylor:2013aa},
we choose $N_{\rm bin}=$~20 and $N_{\rm jk}=$~100 to
determine the covariance matrices of the $C_l$ measurements. This yields
an uncertainty in the error bars of the extracted parameters of 16\%.

Note that when calculating the inverse covariance, we need to multiply it
by the de-biasing factor introduced by
 \cite{Hartlap:2007aa} and \cite{Taylor:2013aa}:
\begin{equation}
f_{\rm corr} = \frac{N_{\rm jk} - 1}{N_{\rm jk} - N_{\rm  bin} - 2} \quad ,
\label{eq:hartlap}
\end{equation}
where $N_{\rm  bin}$ is the size of the data vector, i.e.\ the number
of $C_l$ bins $N_{\rm  bin}$.
In the following section we will present the results of our analysis.

%

\section{Results}
\label{sect:analysis}

In Figure~\ref{cl_data_and_fits} we present the measured angular power
spectra for the galaxy ($gg$),
cluster ($c_{\rm all}c_{\rm all}$) and
galaxy-cluster ($gc_{\rm all}$) cases using the full cluster sample $c_{\rm all}$ and
the analysis method described in Section~\ref{sect:measuring}.
The best fitting models {to these measurements} are
described in Sections~\ref{sect:fitting_nfixed} and~\ref{sect:fitting_n} below
and are shown as lines in Figure~\ref{cl_data_and_fits}.

In order to determine the effective bias for the different tracers, we use two
different models to account for the shot noise. We first analyse the data using
Poissonian shot noise as described in Section~\ref{sect:shot_noise} and
discuss the results of those fits. In a second step we account for
{non-Poissonian} shot noise as discussed in~\ref{sect:effective_n}.

\subsection{Fitting the angular power spectra with a fixed shot noise contribution}
\label{sect:fitting_nfixed}

We first fit to the $C^{\rm data}_l$ the model given by:
\begin{equation}
C_l^{\rm model}(b)
= C_l^{\rm th}(b)
+ {\widetilde N}_l(\bar n) \quad ,
\label{model_fixed_n}
\end{equation}
where $C_l^{\rm th}$ is the theoretical angular power spectrum
as given in Equation~(\ref{eq:class_cl}) for a given value of the effective bias $b$
and
${\widetilde N}_l(\bar n)$ is the Poisson noise term given in
Equation~(\ref{eq:shot_noise_approx}).

For the covariance, there are two contributions: the covariance from the
power spectrum
measurement given in Equation~(\ref{cov_data}), as well as the covariance
from the Poisson noise contribution ${\widetilde N}_l$ given in
Equation~(\ref{eq:neff_cov}).
The covariance of the shot noise contribution
arises from the fact that we determine the shot noise  from
the PolSpice measurements of random maps as discussed in
Section~\ref{sect:shot_noise}.

However, ${\rm Cov}[C^{{\rm rand},i}_l,C^{{\rm rand},i}_m]$ is
about one order of magnitude smaller than ${\rm Cov}^{\rm data}[C_l,C_m]$ and
$N^{-1/2}_s=10$, and therefore the covariance originating from the Poisson noise correction
${\rm Cov}[{\widetilde N}_l,{\widetilde N}_m]$ is about
two orders of magnitude smaller than the covariance contribution from the data ${\rm Cov}^{\rm data}[C_l,C_m]$.
Hence we neglect
the covariance contribution of the shot noise error and use
\begin{equation}
{\rm Cov}^{\rm fit}[C_l,C_m] = {\rm Cov}^{\rm data}[C_l,C_m]
\end{equation}
as the covariance in the Gaussian likelihood ${\mathcal L}$ {of} the effective bias parameters $b$.

We then use this covariance to calculate the Gaussian likelihoods
of the effective bias parameters from all spectra we consider, i.e. from galaxy
and cluster auto-spectra and from the cross-spectrum; we label these
likelihoods  ${\mathcal L}_{g}$,  ${\mathcal L}_{gc_{\rm all}}$, and  ${\mathcal L}_{c_{\rm all}}$ respectively.

Additionally, we can estimate the effective bias likelihood from the cross-correlation
given the results from the two corresponding auto-correlations. For example,
from the likelihoods of the galaxy and cluster auto-correlations
${\mathcal L}_{g}$ and ${\mathcal L}_{c}$, we can construct the following
likelihood
\begin{equation}
{\mathcal L}_{\sqrt{b_gb_c}}(b) = \iint {\rm d} \tilde{b}_g {\rm d} \tilde{b}_c
\, \delta_{\rm D}\left(b - \sqrt{\tilde{b}_g\tilde{b}_c} \right) {\mathcal
  L}_{g}(\tilde{b}_g){\mathcal L}_{c}(\tilde{b}_c) \quad .
\label{eq:likelihood_crosscheck}
\end{equation}
This likelihood serves as a cross-check for the biases obtained by our
analysis and is equivalent to drawing values
for $b_g$ and $b_c$ from the respective auto-correlation distributions and determining
a new distribution from the corresponding $\sqrt{b_gb_c}$.
We determine these cross-check likelihoods and bias values for all the
cross-correlations we determine in our analysis.

\begin{figure}
  \hspace*{-0.5cm}
	\centering \includegraphics[scale=0.45,
          clip=true]{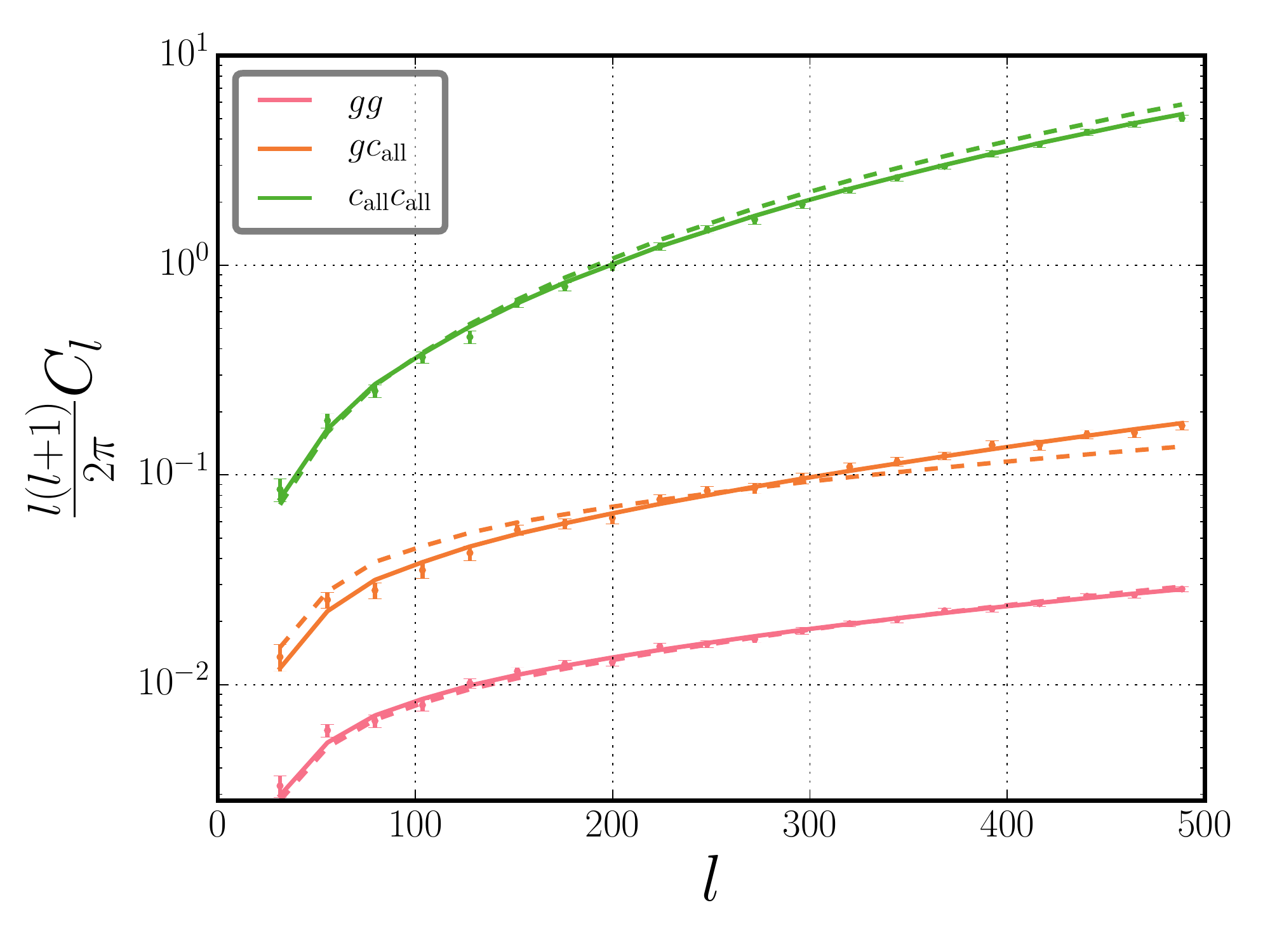}
	\caption{
          Galaxy ($gg$ -- red lines and
          symbols), cluster ($c_{\rm all}c_{\rm all}$ -- green lines and symbols) and galaxy-cluster
          ($gc_{\rm all}$ -- orange lines and symbols) angular power spectra for the data
          described in Section~\ref{sect:data}. Lines indicate the best-fit models
          described in Sections~\ref{sect:fitting_nfixed} and~\ref{sect:fitting_n}:
          dashed lines show the best-fit model specified in Equation~(\ref{model_fixed_n}) using Poissonian
          shot noise, while solid lines use the
          best-fit model specified in Equation~(\ref{model_fit_n}) where $\bar
          n^{\rm eff}$ is added as a fit parameter to adjust the shot noise
          contribution. }
        \label{cl_data_and_fits}
\end{figure}

\begin{figure}
  \hspace*{-1cm}
	\centering \includegraphics[scale=0.45,
          clip=true]{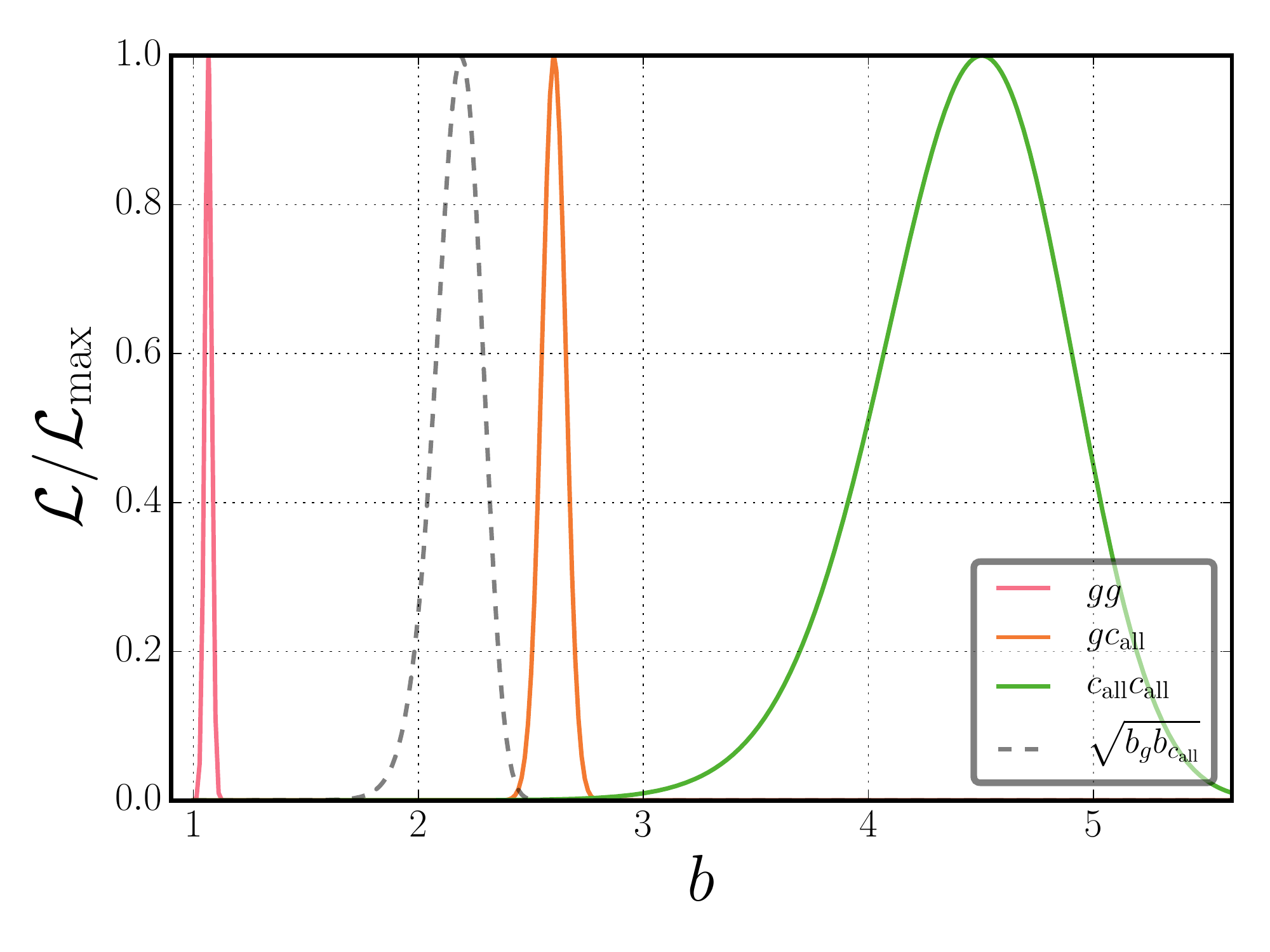}
	\caption{ \label{fig:likelihoods_nfixed}
          Likelihood functions for the effective bias of the samples we consider, obtained using
          a fixed shot noise contribution, for  galaxies ($gg$ -- red lines), clusters
          ($c_{\rm all}c_{\rm all}$ -- green lines) and galaxy-clusters ($gc_{\rm all}$ -- brown lines), as well
          as for the cross-check case ${\mathcal L}_{\sqrt{b_gb_c}}$
          ($\sqrt{b_{g}b_{c}}$ - dashed grey line) described in Section~\ref{sect:fitting_nfixed}.}
\end{figure}

We start by analysing the auto- and cross-correlations for the largest samples available, i.e.\
the galaxy sample $g$ and the full cluster sample $c_{\rm all}$, as we expect to obtain the most accurate
measurements from these samples.
The best-fit models as defined in
Equation~(\ref{model_fixed_n}) are shown as {\em dashed\/} lines in
Figure~\ref{cl_data_and_fits}. While the galaxy auto-correlation ($gg$) is
accurately described by the model, the galaxy-cluster cross-correlation ($gc_{\rm all}$)
as well as the cluster auto-correlation ($cc_{\rm all}$) are poorly described by the
fit. Best-fit parameters, as well as $\chi^2$, are listed in rows 1 -- 3 of
Table~\ref{results_table}. Note that the  $\chi^2$ values are too large
for the $gc_{\rm all}$ and $c_{\rm all}c_{\rm all}$ correlations, indicating that the model is not a
good description of the measurements.
The likelihoods for the fits and cross-check are shown in
Figure~\ref{fig:likelihoods_nfixed} and indicate that also ${\mathcal
  L}_{\sqrt{b_gb_c}}$ does not agree well with the results for ${\mathcal
  L}_{g}$ and ${\mathcal L}_{c}$, i.e.\ the results of the auto- and cross-correlations
are inconsistent with each other.

\begin{table*}
	\centering
	\caption{Results for the different parameter fits from the angular power spectra we use. Column 2 indicates the correlator used,
          column 3 indicates the maximum likelihood value for bias $b$ and the statistical
          error for the fit, column 4 the $\chi^2$ and number of degrees of freedom.
          If the effective noise contribution is also determined in the fit,
          the inverse of the maximum likelihood
          effective pixel density $\bar n^{\rm eff, ML}$ is listed in column 5, while
          column 6 lists the actual inverse pixel density of objects $\bar n^{-1}$
          (only available for auto-correlations). In case
          the cross-check  can
          be performed, the corresponding value for bias $b_{\rm cross-check}$ (including error) is listed in column 7.
        }
	\label{results_table}
	\begin{tabular}{ccccccc}
          \toprule
Row & correlator & $b$ $\pm$ $\sigma_{\rm stat}$ & $\chi^2$/dof & $1/\bar n^{\rm eff} \pm \sigma_{\rm stat}$  & $1/\bar n$ &   $b_{\rm cross-check}$\\ \midrule
1 & $gg$ & 1.07 $\pm$  0.02 & 24.1/19 & --  & 0.026 & --\\
2 & $gc_{\rm all}$ & 2.60 $\pm$ 0.05 & 216/19 & -- & -- & 2.19 $\pm$ 0.10 \\
3 & $c_{\rm all}c_{\rm all}$ & 4.50 $\pm$ 0.42 & 48.9/19 & -- & 30.3 & --\\ \hline
4 & $gg$ & 1.10 $\pm$ 0.03 & 11.8/18 & 0.013 $\pm$ 0.011 & 0.026 & --\\
5 & $gc_{\rm all}$ & 2.29 $\pm$ 0.09 & 18.3/18 & 0.405  $\pm$ 0.086 & -- &  2.30 $\pm$ 0.10 \\
6 & $c_{\rm all}c_{\rm all}$ & 4.82 $\pm$ 0.41 & 14.0/18 & 27.0  $\pm$ 1.90 & 30.3 & --\\ \hline
7 & $gc_{\rm vlim}$ & 2.15 $\pm$ 0.09 & 22.0/18 &  0.610  $\pm$ 0.111 & -- & 2.12 $\pm$ 0.13 \\
8 & $c_{\rm vlim}c_{\rm vlim}$ & 4.09 $\pm$ 0.47 & 8.0/18 &  66.3  $\pm$ 0.067 & 71.4 & --  \\ \hline
9 & $gc_{\lambda_{\text{low}}}$ & 2.11 $\pm$ 0.10 & 17.8/18 &  0.314  $\pm$ 0.107 & -- & 2.09 $\pm$ 0.13\\
10 & $gc_{\lambda_{\text{high}}}$ & 2.53 $\pm$ 0.13 & 9.43/18 &  0.516  $\pm$ 0.119 & -- & 2.50 $\pm$ 0.15 \\
11 & $c_{\lambda_{\text{low}}}c_{\lambda_{\text{low}}}$ & 3.99 $\pm$ 0.48 & 13.3/18 &  57.0  $\pm$ 3.64 & 62.5 & --\\
12 & $c_{\lambda_{\text{high}}}c_{\lambda_{\text{high}}}$ & 5.72 $\pm$ 0.65 & 9.48/18 &  57.6  $\pm$ 4.14 & 62.5 & --\\
13 & $c_{\lambda_{\text{low}}}c_{\lambda_{\text{high}}}$ & 4.98 $\pm$ 0.57 & 11.0/18 &  -3.69  $\pm$ 1.60 & -- & 4.71 $\pm$ 0.41 \\
          \bottomrule
\end{tabular}
\end{table*}

\begin{figure}
  \hspace*{-0.5cm}
	\centering \includegraphics[scale=0.45,
          clip=true]{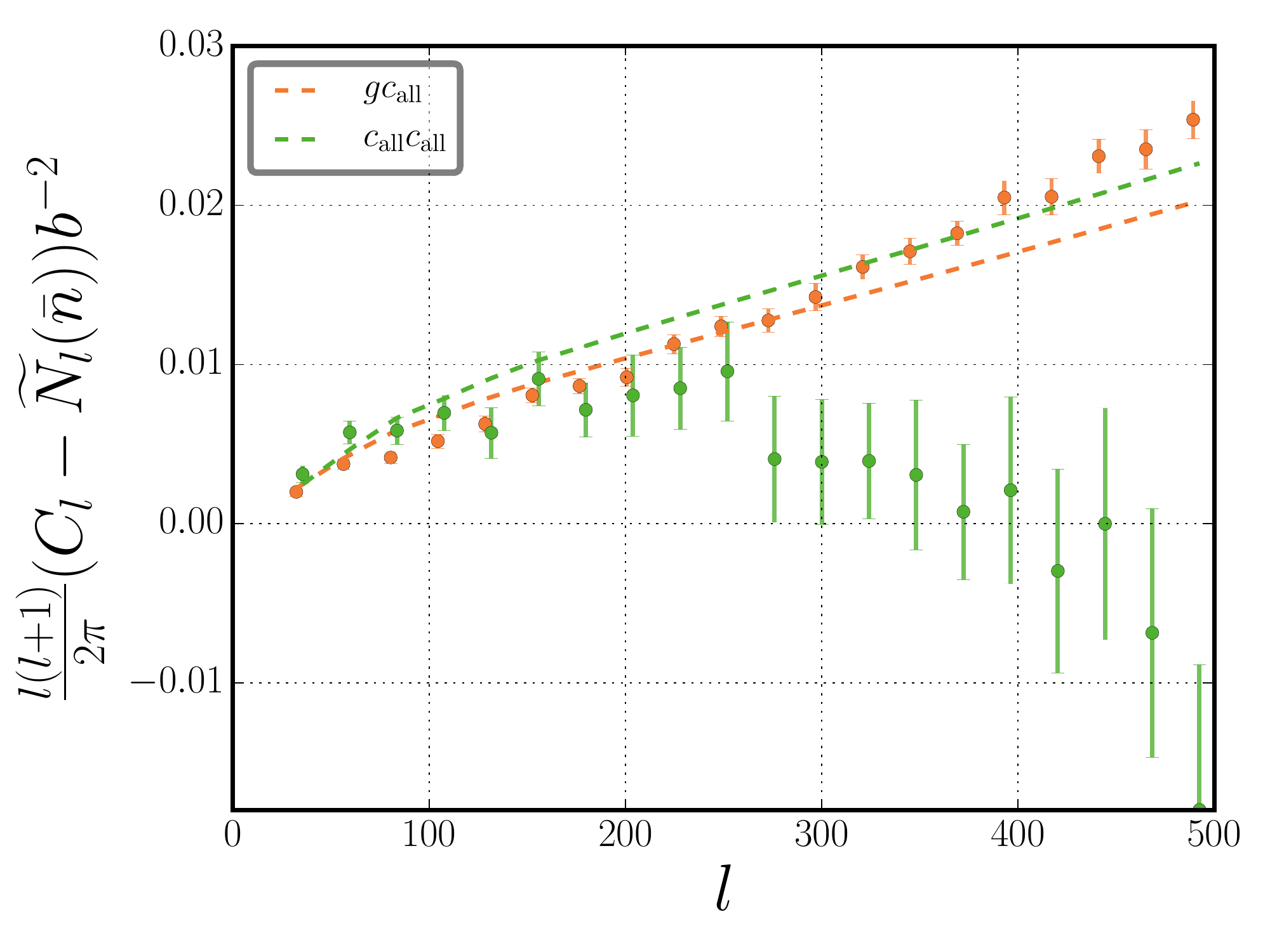}
	\caption{ \label{fig:noise_removed_fit}
          Cluster ($c_{\rm all}c_{\rm all}$) and galaxy-cluster
          ($gc_{\rm all}$) power spectra with Poissonian shot noise ${\widetilde N}_l$
          removed according to
          Equation~(\ref{eq:noise_const}) and rescaled by the best-fit bias $b$.
          Dashed lines show the best-fit model specified in Equation~(\ref{model_fixed_n}).}
 \end{figure}

Concentrating on the mismatch between the data and model for the
galaxy-cluster ($gc_{\rm all}$) and cluster ($c_{\rm all}c_{\rm all}$) correlation, we remove the shot noise
contribution, which is dominant for the $c_{\rm all}c_{\rm all}$ correlation, from both the data and
model. In addition, we bring the $C_l$ to the same scale by renormalizing them
by the best fit bias; the resulting $C_l$ are shown in
Figure~\ref{fig:noise_removed_fit}.

For the galaxy-cluster
cross-correlation, the model (orange symbols and line)
clearly overestimates the $C_l$ for low $l$ and
underestimates them for high $l$. This is not unexpected, because neglecting the shot
noise contribution for the galaxies that are part of clusters (as discussed
in Section~\ref{sect:shot_noise}) is expected to be a poor approximation.

In the case of the cluster auto-correlation, we can now see the mismatch between the data and model
more clearly because the extracted signal is dominated by the shot noise.
Also in this case, the figure shows a severe mismatch,
as the model overestimates the  $C_l$ on all scales. Above $l\approx400$ the noise-corrected
$C_l$ even turn negative, indicating the shot noise is overcorrected
using the average object per pixel density $\bar n$ for clusters.

\subsection{Accounting for the effective noise contribution}
\label{sect:fitting_n}

Following the reasoning of Section~\ref{sect:effective_n},
we can effectively account for a modification of the shot noise contribution
if we limit ourselves to a regime where the correction is
(to a good approximation) independent of $l$. We can then introduce
an effective number density of objects per pixel
$\bar n^{\rm eff}$ as a nuisance parameter.
This allows us to account for any sub- and super-Poissonian shot noise
contributions where the shape of the shot noise correction is unaffected
and only the
magnitude of $\widetilde{N}_l$ is adjusted, i.e.
\begin{equation}
C_l^{\rm model}(b, \bar n^{\rm eff} ) = C_l^{\rm th}(b)
+ {\widetilde N}_l(\bar n^{\rm eff} ) \quad .
\label{model_fit_n}
\end{equation}

To account for measuring systematics affecting the power spectra, a similar
treatment was used
by~\citet{Ho:2012aa,Zhao:2013aa,Beutler:2014aa,Johnson:2016aa,Grieb:2016aa}
in their analyses.

\begin{figure}
	\centering
    \hspace*{-0.5cm}
  \includegraphics[scale=0.45,
          clip=true]{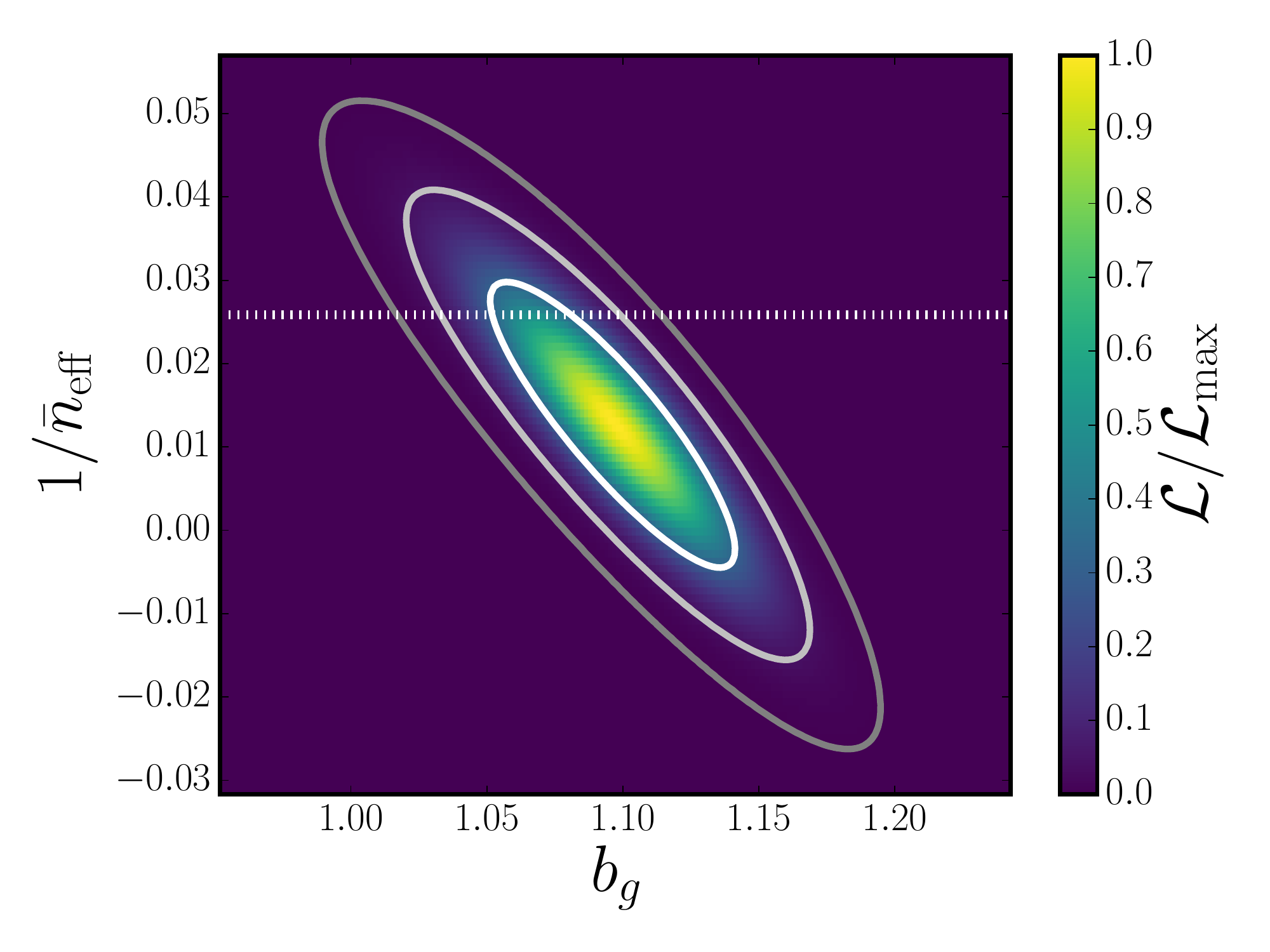}
          \hspace*{-0.5cm}
        \includegraphics[scale=0.45,
          clip=true]{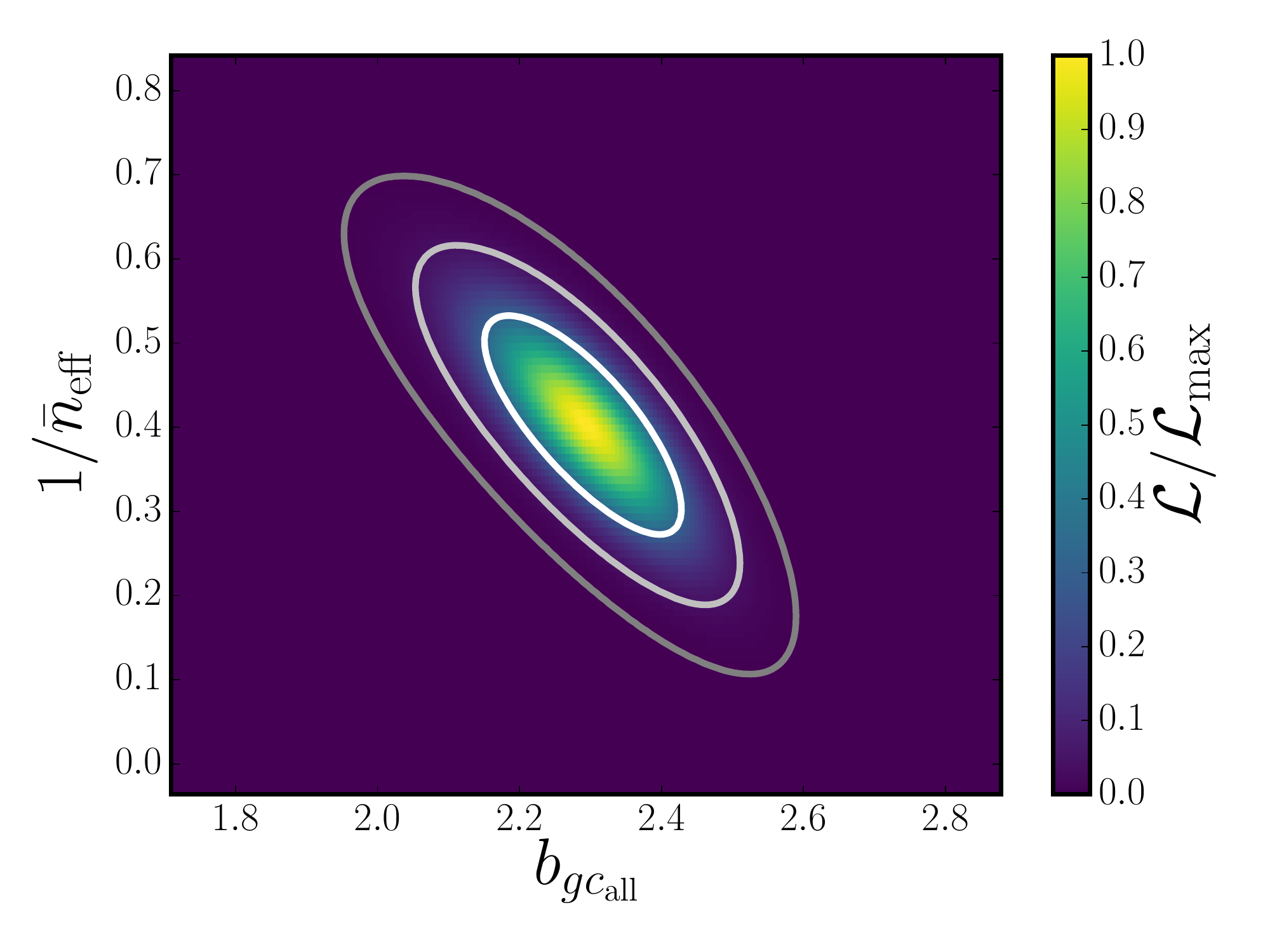}
          \hspace*{-0.5cm}
        \includegraphics[scale=0.45,
          clip=true]{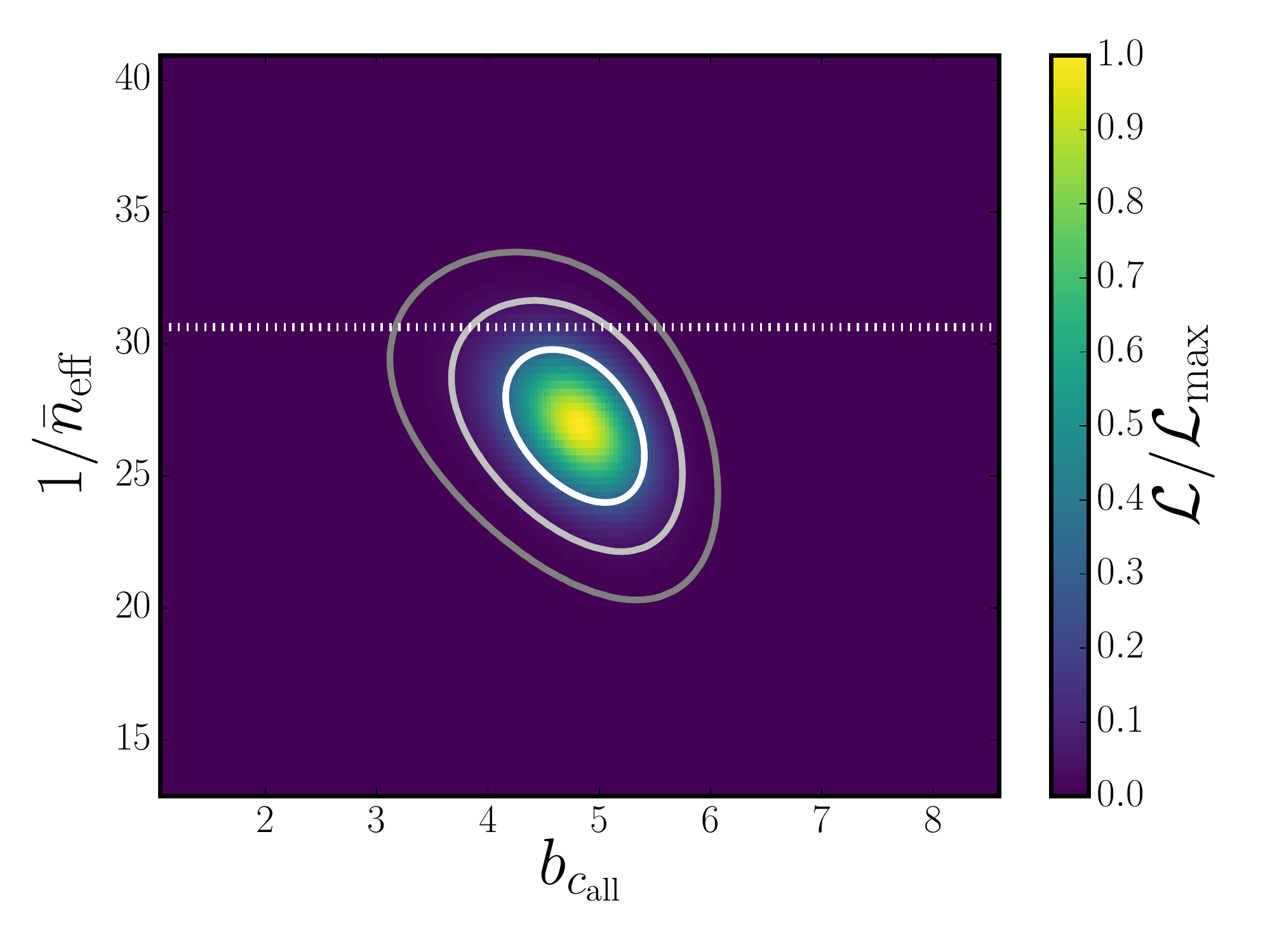}
	\caption{ \label{fig:likelihoods_fitn}
          The {2d}-likelihoods for the fits
          described in Section~\ref{sect:fitting_n} for the parameters bias $b$ and
          effective number of objects per pixel $\bar n^{\rm eff}$.
          Solid lines indicate the $68\%$, $95\%$ and $99\%$ confidence regions respectively.  For
          the auto-correlation cases the dotted lines indicate the actual
          inverse number of objects per pixel $\bar n^{-1}$. }
%
\end{figure}

The results for the fits including the effective shot noise contribution
as a free parameter are shown as the {\em solid} lines in
Figure~\ref{cl_data_and_fits}. The $C^{\rm model}_l(b, \bar n^{\rm eff} )$ for the fit describe the data
much more accurately. The corresponding constant likelihood contours are shown in
Figure~\ref{fig:likelihoods_fitn} where the dashed lines indicate the actual $1/\bar n$
for the Poissonian shot noise contribution (not applicable to the cross-correlation $gc_{\rm all}$).
Both the galaxy and cluster auto-correlations
slightly favour sub-Poissonian noise contributions, though the deviation
from the Poissonian noise case is below 1-$\sigma$ for galaxies and about 1.5-$\sigma$
for the clusters. As expected, the noise contribution for the
galaxy auto-correlation is small. However, the shot noise for the cluster
auto-correlation is a dominant contribution, and therefore we see a much better description
of the data compared to the previous fit where we only fitted for the bias. The same
holds for the galaxy-cluster cross-correlation for which the data clearly favours
a non-zero shot noise contribution.

\begin{figure}
  \hspace*{-1cm}
	\centering \includegraphics[scale=0.45,
          clip=true]{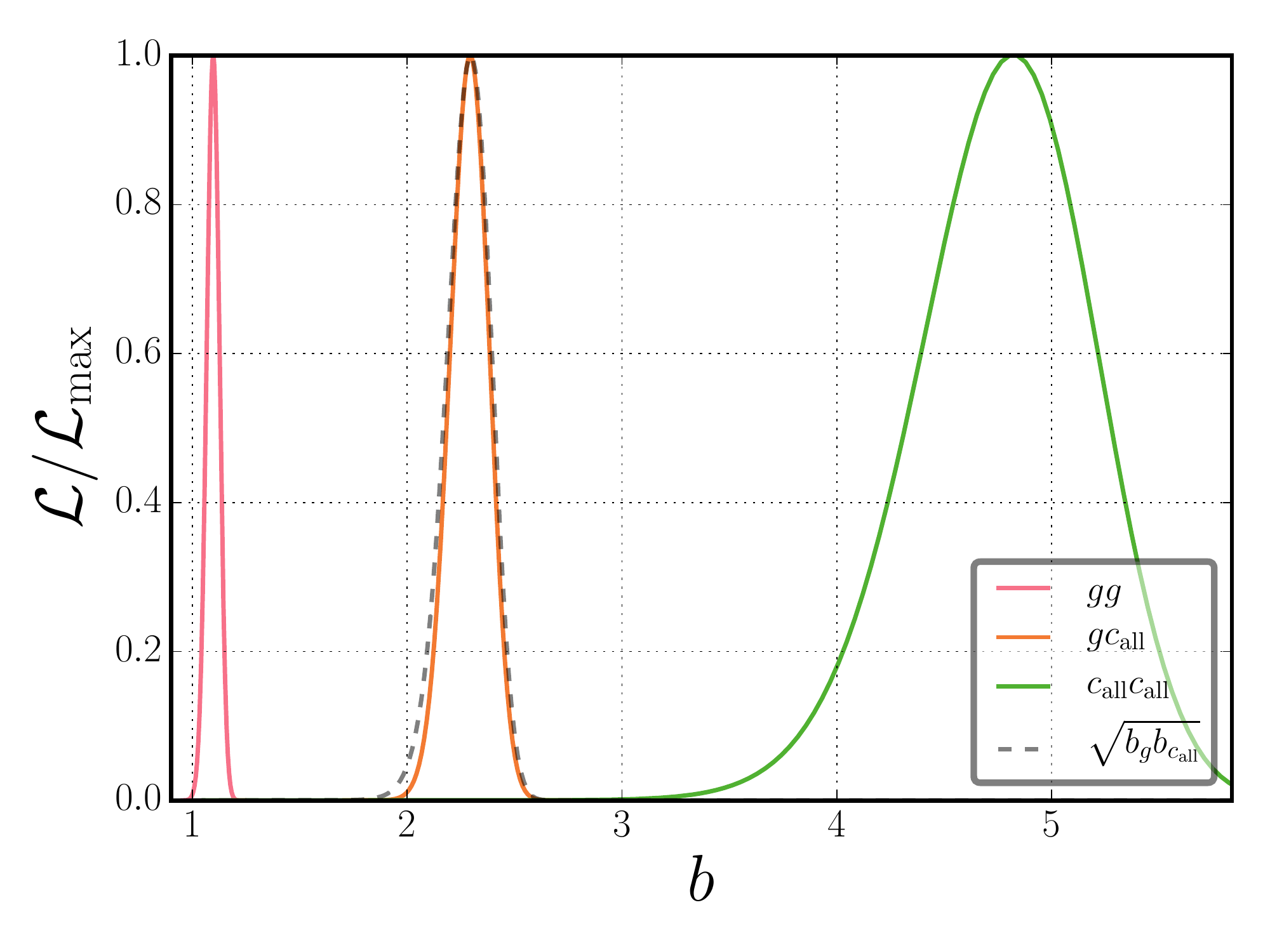}
	\caption{ \label{fig:marginalised_likelihoods_fitting_n}
          Likelihoods of the effective bias, marginalised over the {amplitude} of the shot
        noise contribution as described in Section~\ref{sect:fitting_n}, for the different correlators we consider: galaxies auto- ($gg$ -- red lines),
          clusters auto- ($c_{\rm all}c_{\rm all}$ -- green lines) and galaxy-cluster cross-correlation ($gc_{\rm all}$ -- brown lines) as well as the consistency
          check bias ($\sqrt{b_{g}b_{c}}$ -- dashed
          grey line) as described in Section~\ref{sect:fitting_nfixed}. }  

\end{figure}

We show in
Figure~\ref{fig:marginalised_likelihoods_fitting_n} the marginalised likelihoods for the bias parameters;
we can see that in this case the cross-check likelihood ${\mathcal L}_{\sqrt{b_gb_c}}$ agrees well with the
result obtained from the galaxy-cluster cross-correlation $gc_{\rm all}$. This means that
the measurements of auto- and cross-correlation are {now} in good agreement with each other
when introducing $n_{\rm eff}$ as an additional model parameter.

The results of these fits, including maximum likelihood values for
${\mathcal   L}_{\sqrt{b_gb_c}}$, are summarised in rows 4-6 of
Table~\ref{results_table}. Note that
the $\chi^2$ have improved significantly compared to the bias-only fits.

The fact that we obtain a sub-Poissonian shot noise from the auto-correlation
of galaxies argues for a relatively low satellite fraction in the sample,
as discussed in Section~\ref{sec:non_poissonian_shot_noise}.
The sub-Poissonian shot noise obtained from the cluster
auto-correlation is expected because clusters obey halo exclusion.

\subsection{Effective bias for volume-limited cluster sample}
\label{sect:vlim}

The measurement of the cluster auto-correlation for the full sample $c_{\rm all}$ yields
a bias value of $b_{c} = 4.82 \pm 0.41 $.
However, the $c_{\rm all}$ sample is not volume-limited and we should therefore
use the volume-limited sample $c_{\rm vlim}$ when comparing the effective bias
to theoretical expectations.

We perform the same analysis of Section~\ref{sect:fitting_n} on the volume-limited
sample $c_{\rm vlim}$ as discussed in Section~\ref{sect:data} and
summarized in Table~\ref{tab:samples}. The fits for the volume-limited cluster
sample are qualitatively similar to the $c_{\rm all}$ samples and we therefore do not
show plots of the {power spectra} and best fits. Results for the
auto-correlation $c_{\rm vlim}c_{\rm vlim}$ as well
as the cross-correlation $gc_{\rm vlim}$ are listed in rows 7 and 8 of
Table~\ref{results_table}. The bias from the cross-check and the bias for the cross-correlation
are again in agreement. {From} the cluster auto-correlation we
find $b_{c_{\rm vlim}} = 4.09 \pm 0.47$.

From the theoretical modeling of the effective bias in Section~\ref{sect:theory_bias}
we expect  $b \approx 2.8$ for the volume-limited
cluster sample, which is in tension with the value extracted from the data
at the 2-3~$\sigma$ level.

\subsubsection{Systematics affecting the bias measurement}
We explored -- and ruled out -- the following systematic effects which might
explain this discrepancy.
The redshift and richness distributions of clusters we expect from the predictions
in Section~\ref{sect:theory_bias}
are in good agreement with those measured for the volume-limited sample as shown
in Figure~\ref{zdist} and cannot be used {to} explain this discrepancy.
Neither are there extremely high mass/bias objects that could explain the difference.
We have checked if statistical and systematic uncertainty of the mass-richness
relation for galaxy clusters could account for the tension. Even shifting the
mass-richness relation by 30\% in mass does not alleviate the tension. The
effect of the measurement uncertainties of the mass-richness relation parameters
is negligible.
The measurement error on $\sigma_8$ from the Planck2013+WP+highL+BAO measurement
also cannot account for the discrepancy, i.e. shifting the value of $\sigma_8$
by 1-$\sigma$ does only have a very small effect on our measurement of the
effective bias.

In this work we have not explored the effect of assembly bias
\citep{2004MNRAS.350.1385S,2005MNRAS.363L..66G,2007MNRAS.377L...5G,2006ApJ...652...71W,More:2016aa},
i.e.\ the dependence of halo clustering on assembly history.
However, our effective bias measurement is consistent with the results
examining this effect
as reported in~\cite{Miyatake:2016aa} and~\cite{Baxter:2016aa} and references therein.
\cite{Miyatake:2016aa} split the clusters into two subsamples based
on the average member galaxy separation from the cluster center and find
significantly different values for the bias for those subsamples: $b=2.17\pm0.31$
for clusters with a low average member galaxy separation and $b=3.67\pm0.40$
for large average separation. \cite{Baxter:2016aa} measures the angular
correlation function $w(\theta)$ for clusters in different richness and
redshift bins and find high (compared to the prediction using the Tinker
mass function) bias values between 3 and 5 for the $\lambda > 20$ richness
bins.
It will be very interesting to study
this effect in more detail using angular power spectra in harmonic space
in a future analysis.

%

\subsection{Results for low and high richness cluster samples}
\label{sect:richness_fits}
Next, we investigate the shot noise properties in the cross-correlations
of clusters in different richness bins and hence different halo mass.
Therefore, we divide the $c_{\rm all}$ sample into two halves, splitting it at the median richness
$\lambda_{\rm median} = 33.7$  with
$\lambda_{\rm low} <  \lambda_{\rm median}$ and $\lambda_{\rm low} > \lambda_{\rm median}$
with median redshifts 0.315 and 0.42. The redshift distributions for
these two sub-samples are shown in Figure~\ref{zdist}.

The auto-correlations of the richness{-split}  cluster samples and their
relative best-fit bias values are qualitatively {similar to}
the $c_{\rm all}$ sample.
We therefore do not show the results for the auto-correlations
and will only discuss and show the cross-correlation measurements below.
{The fit results} are summarized in rows 9-13 of Table~\ref{results_table}. As expected,
for the auto-correlation of  the low-richness sample  $c_{\lambda_{\rm low}}$ we observe
a smaller bias than for the $c_{\rm all}$ sample, while for the high-richness sample
$c_{\lambda_{\rm high}}$ the bias is shifted
to even higher values (rows 11 and 12). The effective shot noise contribution
is larger for these samples as reflected by the smaller values {of} $\bar n_{\rm eff} $,
because there are half as many objects in the sub-samples.
A similar systematic bias shift can be seen for the galaxy-cluster cross-correlation,
which is shown in {rows} 7-8, while the shot-noise contribution is smaller for the
$gc_{\lambda_{\text{low}}}$ and higher for $gc_{\lambda_{\text{high}}}$ cross-correlation.

{From} the cross-correlation {between} the low- and high-richness samples, we expect
a small {or} vanishing value {of} the effective objects per pixel density,
because there is no overlap of objects between
the two samples (except for systematic effects from cluster finding/identification
and line of sight effects). However, the actual value of the effective pixel density is negative
at the 2-$\sigma$ level.
Again, this argues for strong exclusion effects between clusters of different
richness (and thus halo mass), as {observed in} $N$-body
simulations~\citep{Hamaus:2010aa, Baldauf:2013aa}.

\begin{figure}
  \hspace*{-1cm}
	\centering \includegraphics[scale=0.45,
          clip=true]{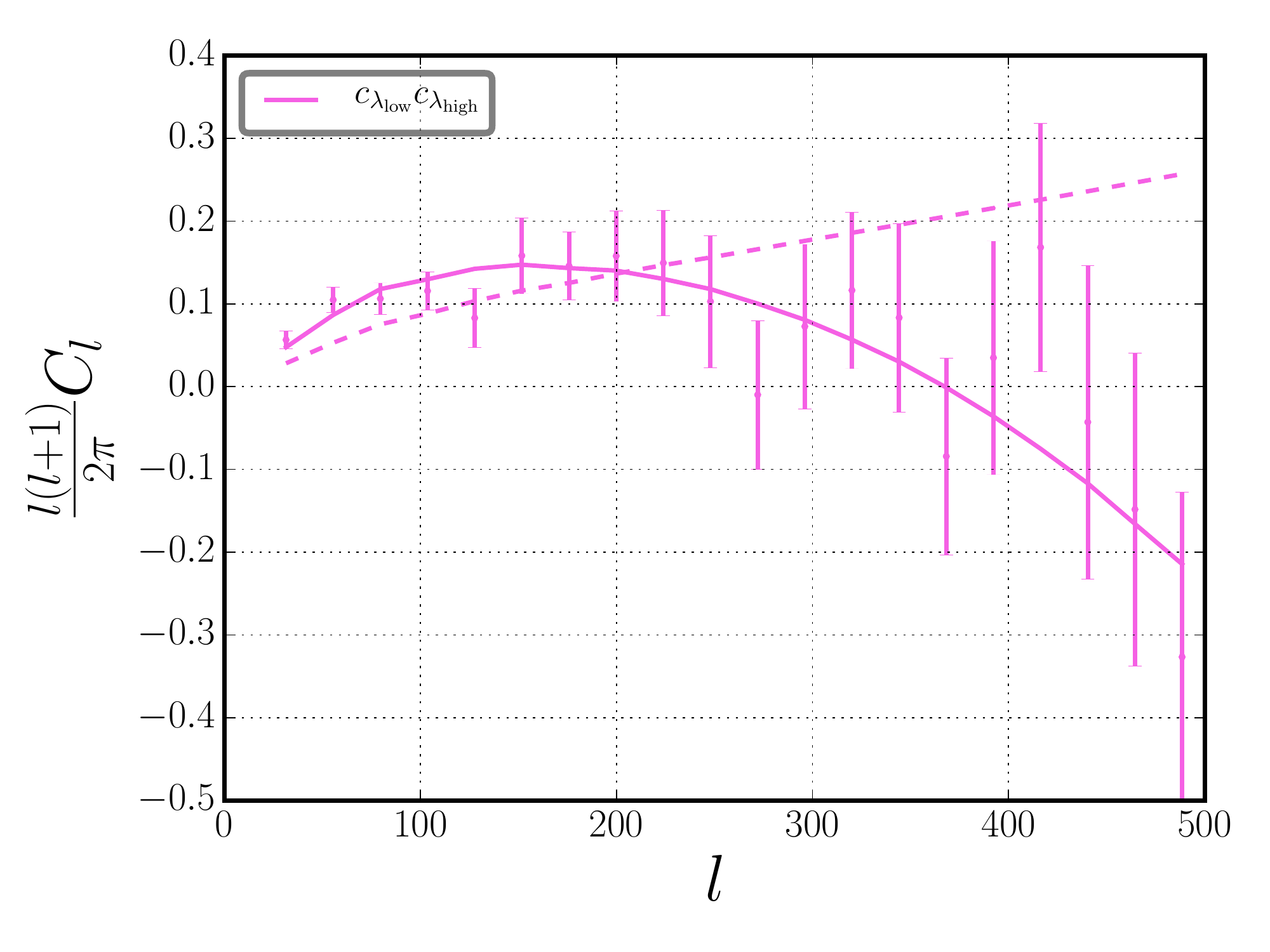}
	\caption{ \label{cl_data_and_fits_low_high}
          Cross-correlation {between} the low- and high-richness samples
          described in Section~\ref{sect:data}. Lines indicate the best-fit models
          described in Sections~\ref{sect:fitting_nfixed} and~\ref{sect:fitting_n}:
          Dashed line shows the best-fit model specified in Equation~(\ref{model_fixed_n}), using a vanishing
          shot-noise contribution (as there is no overlap between the two
          samples), solid line represents the
          best-fit model specified in Equation~(\ref{model_fit_n}), where $\bar
          n^{\rm eff}$ is added as a fit parameter to adjust the shot-noise
          contribution. The best fit $\bar n^{\rm eff}$ is negative in this case.}
\end{figure}

Figure~\ref{cl_data_and_fits_low_high} shows the measured angular power spectrum as well as the
best-fit models using a vanishing shot-noise contribution (dashed line) and fitting
for the shot-noise contribution (solid line). The measured $C_l$ show an unusual
behavior as the correlation increases up to $l\approx200$ and then decreases
again. This cannot be described by a model with a vanishing (shown by the
dashed line) shot-noise contribution, as a positive shot-noise contribution
would worsen this mismatch. However, if allowing for negative
non-Poissonian shot noise, the model yields a good description of the data.
It will be interesting {to see whether} this measurement persists for larger sample sizes and other
data sets.

\begin{figure}
  \hspace*{-0.5cm}
	\centering \includegraphics[scale=0.6,
          clip=true]{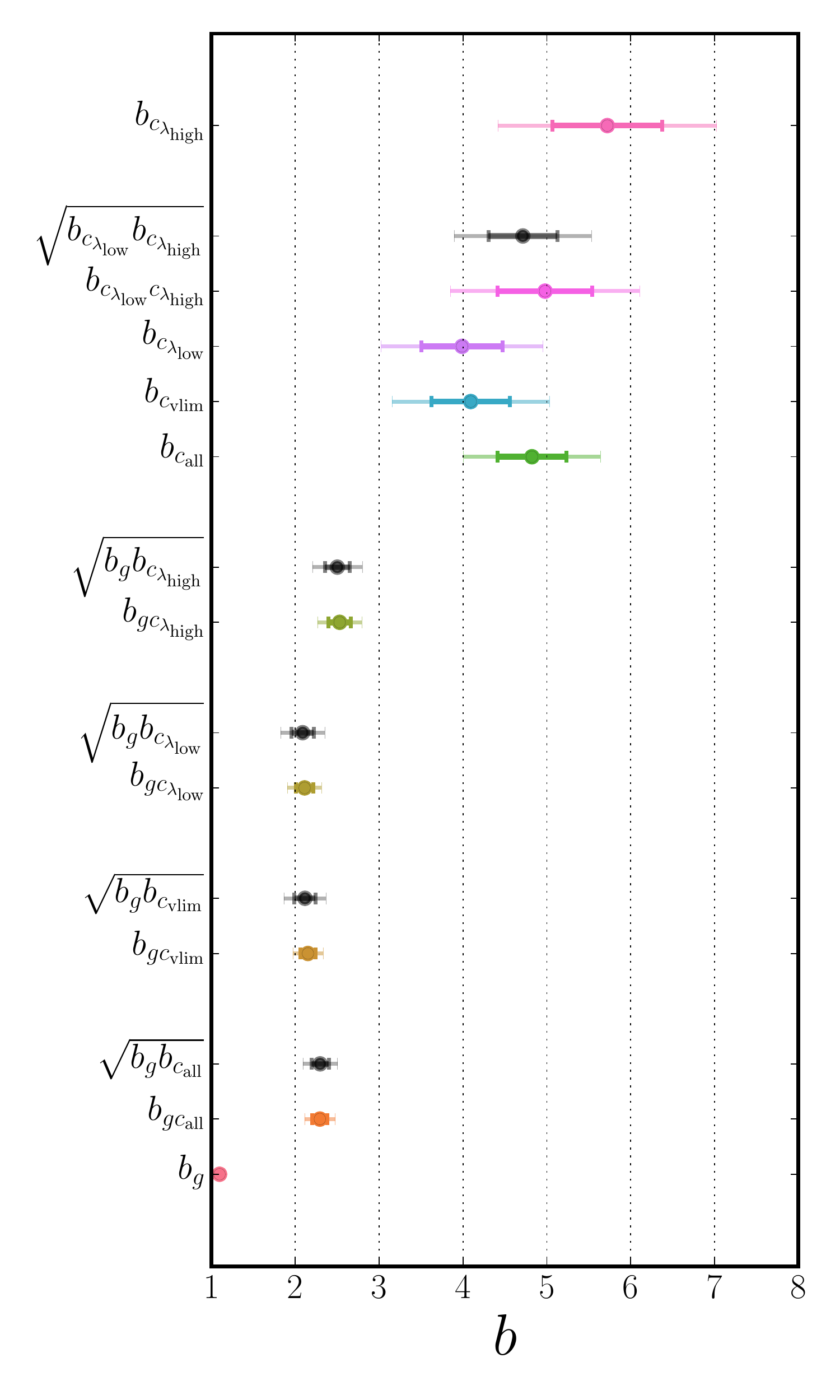}
	\caption{ \label{fig:bias_summary}
          The maximum likelihood galaxy and dark mater halo bias after marginalising over
          over $\bar n^{\rm eff} $,
          as well as the expectations denoted by $\sqrt{b_{a}b_{b}}$ from the cross-check described in Section~\ref{sect:fitting_n}.
          We also show $68\%$ (solid lines) and  $95\%$ (faint solid lines) error bars.
        }
\end{figure}

We have summarized the bias fit results for the different catalogue samples
in Figure~\ref{fig:bias_summary}, which includes 1- and 2-sigma error bars
for all measurements. This figure illustrates that the cross-check
values (black symbols) and errors for bias from Equation~(\ref{eq:likelihood_crosscheck})
of the measurements of auto- and cross-correlations are consistent with
each other for all measurements, including the volume-limited sample $c_{\rm vlim}$
as well as for the two richness bin samples $c_{\lambda_{\rm low}}$ and  $c_{\lambda_{\rm high}}$.

\section{Summary and Outlook}
\label{sect:conclusions}
We have presented a first measurement
of the cross-correlation
angular power spectrum of galaxies and galaxy clusters using the SDSS DR8
galaxy and galaxy cluster sample. Further, we measured the
auto- and cross-correlations of different sub-samples of the full cluster catalogue:
a volume-limited sample as well as two samples of low and high richness.

We argued that in order to get a good theoretical description for
the cross-correlation measurements we need to add an effective shot-noise
contribution as an additional component to our model. Because there is some
overlap of galaxies and galaxy clusters, we expect a non-vanishing shot-noise contribution.
We find the measurements
are much better described by a model containing sub-Poissonian shot noise and
that using a regular Poisson shot noise correction results in an
overcorrection.
Since we also expect a deviation from Poissonian
shot noise for cluster auto-correlations due to halo exclusion and non-linear
clustering~\citep{Baldauf:2013aa}, we  investigated if the cluster auto-correlation
shows deviation from Poissonian shot noise
as expected from simulations~\citep{Hamaus:2010aa}.

We extracted the effective bias for our measurements and used
the results for the effective bias from auto-correlation measurements to perform
a cross-check on the effective bias from the cross-correlation measurements.
These cross-checks were in very good agreement after we allowed for non-Poissonian
shot noise contribution.
We performed the same cross-checks for measurements involving the subsamples of
the cluster data and we find all measurements of effective bias to be consistent.

To compare our measurement of effective bias to theoretical expectations,
we constructed a volume-limited cluster sample and found a relatively
high value of $4.09 \pm 0.47$  compared to our expectation of $b=2.8$.
However, this value
is consistent with previous measurements and supports the case for a fuller
exploration of additional systematic effects, such as halo assembly bias.

Finally, we constructed a low and high richness sample from the full
cluster sample and measured the auto- and cross-correlation as well. Again, the
values for the effective bias are consistent with each other and are all
relatively large for the cluster samples.

Most notably, we found a negative shot noise contribution for the
cross-correlation at the 2-$\sigma$ level. This argues for strong exclusion
effects between clusters of different richness (and thus halo mass), in
agreement with $N$-body simulations.
As larger and better data sets will become available, it will be interesting
to see if this measurement of negative shot noise persists.

An appropriate treatment of shot noise is important for many other auto-
and cross-correlation large scale structure analyses, i.e. different
galaxy types and multi-tracer surveys. In the future, it will be essential to account
for these effects to derive unbiased cosmology constraints from correlation
functions and power spectra analyses.

\section*{Acknowledgments}
\label{ack}
The authors would like to thank O.~Friedrich and M.\ M.~Rau
for fruitful discussions.
M. Costanzi, B. Hoyle and J. Weller acknowledge support from the Trans-Regional
Collaborative Research Center TRR 33 ``The Dark Universe'' of the Deutsche Forschungsgemeinschaft
(DFG).

Some of the results in this paper have been derived using the HEALPix
(\cite{Gorski:2004by}) package.

Funding for SDSS-III has been provided by the Alfred P. Sloan Foundation, the
Participating Institutions, the National Science Foundation, and the
U.S. Department of Energy Office of Science. The SDSS-III web site is
\mbox{http://www.sdss3.org/}.

SDSS-III is managed by the Astrophysical Research Consortium for the
Participating Institutions of the SDSS-III Collaboration including the
University of Arizona, the Brazilian Participation Group, Brookhaven National
Laboratory, Carnegie Mellon University, University of Florida, the French
Participation Group, the German Participation Group, Harvard University, the
Instituto de Astrofisica de Canarias, the Michigan State/Notre Dame/JINA
Participation Group, Johns Hopkins University, Lawrence Berkeley National
Laboratory, Max Planck Institute for Astrophysics, Max Planck Institute for
Extraterrestrial Physics, New Mexico State University, New York University, Ohio
State University, Pennsylvania State University, University of Portsmouth,
Princeton University, the Spanish Participation Group, University of Tokyo,
University of Utah, Vanderbilt University, University of Virginia, University of
Washington, and Yale University.

\bibliographystyle{mnras}
\bibliography{literature}

\begin{thebibliography}{}
\makeatletter
\relax
\def\mn@urlcharsother{\let\do\@makeother \do\$\do\&\do\#\do\^\do\_\do\%\do\~}
\def\mn@doi{\begingroup\mn@urlcharsother \@ifnextchar [ {\mn@doi@}
  {\mn@doi@[]}}
\def\mn@doi@[#1]#2{\def\@tempa{#1}\ifx\@tempa\@empty \href
  {http://dx.doi.org/#2} {doi:#2}\else \href {http://dx.doi.org/#2} {#1}\fi
  \endgroup}
\def\mn@eprint#1#2{\mn@eprint@#1:#2::\@nil}
\def\mn@eprint@arXiv#1{\href {http://arxiv.org/abs/#1} {{\tt arXiv:#1}}}
\def\mn@eprint@dblp#1{\href {http://dblp.uni-trier.de/rec/bibtex/#1.xml}
  {dblp:#1}}
\def\mn@eprint@#1:#2:#3:#4\@nil{\def\@tempa {#1}\def\@tempb {#2}\def\@tempc
  {#3}\ifx \@tempc \@empty \let \@tempc \@tempb \let \@tempb \@tempa \fi \ifx
  \@tempb \@empty \def\@tempb {arXiv}\fi \@ifundefined
  {mn@eprint@\@tempb}{\@tempb:\@tempc}{\expandafter \expandafter \csname
  mn@eprint@\@tempb\endcsname \expandafter{\@tempc}}}

\bibitem[\protect\citeauthoryear{{Abell}}{{Abell}}{1958}]{Abell1958}
{Abell} G.~O.,  1958, \mn@doi [Astrophys.J.Suppl.] {10.1086/190036}, \href
  {http://adsabs.harvard.edu/abs/1958ApJS....3..211A} {3, 211}

\bibitem[\protect\citeauthoryear{Aihara et~al.}{Aihara
  et~al.}{2011}]{Aihara:2011sj}
Aihara H.,  et~al., 2011, \mn@doi [Astrophys.J.Suppl.]
  {10.1088/0067-0049/193/2/29}, 193, 29

\bibitem[\protect\citeauthoryear{{Anderson} et~al.,}{{Anderson}
  et~al.}{2012}]{Anderson:2012aa}
{Anderson} L.,  et~al., 2012, \mn@doi [\mnras]
  {10.1111/j.1365-2966.2012.22066.x}, \href
  {http://adsabs.harvard.edu/abs/2012MNRAS.427.3435A} {427, 3435}

\bibitem[\protect\citeauthoryear{{Balaguera-Antol{\'{\i}}nez}, {S{\'a}nchez},
  {B{\"o}hringer}, {Collins}, {Guzzo}  \&
  {Phleps}}{{Balaguera-Antol{\'{\i}}nez}
  et~al.}{2011}]{Balaguera-Antolinez:2011aa}
{Balaguera-Antol{\'{\i}}nez} A.,  {S{\'a}nchez} A.~G.,  {B{\"o}hringer} H.,
  {Collins} C.,  {Guzzo} L.,   {Phleps} S.,  2011, \mn@doi [\mnras]
  {10.1111/j.1365-2966.2010.18143.x}, \href
  {http://adsabs.harvard.edu/abs/2011MNRAS.413..386B} {413, 386}

\bibitem[\protect\citeauthoryear{{Baldauf}, {Seljak}, {Smith}, {Hamaus}  \&
  {Desjacques}}{{Baldauf} et~al.}{2013}]{Baldauf:2013aa}
{Baldauf} T.,  {Seljak} U.,  {Smith} R.~E.,  {Hamaus} N.,   {Desjacques} V.,
  2013, \mn@doi [\prd] {10.1103/PhysRevD.88.083507}, \href
  {http://adsabs.harvard.edu/abs/2013PhRvD..88h3507B} {88, 083507}

\bibitem[\protect\citeauthoryear{{Baxter}, {Rozo}, {Jain}, {Rykoff}  \&
  {Wechsler}}{{Baxter} et~al.}{2016}]{Baxter:2016aa}
{Baxter} E.~J.,  {Rozo} E.,  {Jain} B.,  {Rykoff} E.,   {Wechsler} R.~H.,
  2016, preprint, \href {http://adsabs.harvard.edu/abs/2016arXiv160400048B} {}
  (\mn@eprint {arXiv} {1604.00048})

\bibitem[\protect\citeauthoryear{{Bennett} et~al.,}{{Bennett}
  et~al.}{2013}]{Bennett:2013aa}
{Bennett} C.~L.,  et~al., 2013, \mn@doi [\apjs] {10.1088/0067-0049/208/2/20},
  \href {http://adsabs.harvard.edu/abs/2013ApJS..208...20B} {208, 20}

\bibitem[\protect\citeauthoryear{{Beutler} et~al.,}{{Beutler}
  et~al.}{2011}]{Beutler:2011aa}
{Beutler} F.,  et~al., 2011, \mn@doi [\mnras]
  {10.1111/j.1365-2966.2011.19250.x}, \href
  {http://adsabs.harvard.edu/abs/2011MNRAS.416.3017B} {416, 3017}

\bibitem[\protect\citeauthoryear{{Beutler} et~al.,}{{Beutler}
  et~al.}{2014}]{Beutler:2014aa}
{Beutler} F.,  et~al., 2014, \mn@doi [\mnras] {10.1093/mnras/stu1051}, \href
  {http://adsabs.harvard.edu/abs/2014MNRAS.443.1065B} {443, 1065}

\bibitem[\protect\citeauthoryear{{Blake} et~al.,}{{Blake}
  et~al.}{2011}]{Blake:2011aa}
{Blake} C.,  et~al., 2011, \mn@doi [\mnras] {10.1111/j.1365-2966.2011.19592.x},
  \href {http://adsabs.harvard.edu/abs/2011MNRAS.418.1707B} {418, 1707}

\bibitem[\protect\citeauthoryear{{Blas}, {Lesgourgues}  \& {Tram}}{{Blas}
  et~al.}{2011}]{Blas:2011aa}
{Blas} D.,  {Lesgourgues} J.,   {Tram} T.,  2011, \mn@doi [\jcap]
  {10.1088/1475-7516/2011/07/034}, \href
  {http://adsabs.harvard.edu/abs/2011JCAP...07..034B} {7, 034}

\bibitem[\protect\citeauthoryear{{Challinor} \& {Chon}}{{Challinor} \&
  {Chon}}{2005}]{Challinor:2005aa}
{Challinor} A.,  {Chon} G.,  2005, \mn@doi [\mnras]
  {10.1111/j.1365-2966.2005.09076.x}, \href
  {http://adsabs.harvard.edu/abs/2005MNRAS.360..509C} {360, 509}

\bibitem[\protect\citeauthoryear{Chon, Challinor, Prunet, Hivon  \&
  Szapudi}{Chon et~al.}{2004}]{Chon:2003gx}
Chon G.,  Challinor A.,  Prunet S.,  Hivon E.,   Szapudi I.,  2004, \mn@doi
  [Mon.Not.Roy.Astron.Soc.] {10.1111/j.1365-2966.2004.07737.x}, 350, 914

\bibitem[\protect\citeauthoryear{Cole et~al.}{Cole et~al.}{2005}]{Cole:2005sx}
Cole S.,  et~al., 2005, \mn@doi [Mon.Not.Roy.Astron.Soc.]
  {10.1111/j.1365-2966.2005.09318.x}, 362, 505

\bibitem[\protect\citeauthoryear{Collins, Guzzo, Boehringer, Schuecker,
  Chincarini  et~al.}{Collins et~al.}{2000}]{Collins:2000yk}
Collins C.,  Guzzo L.,  Boehringer H.,  Schuecker P.,  Chincarini G.,   et~al.,
  2000, \mn@doi [Mon.Not.Roy.Astron.Soc.] {10.1046/j.1365-8711.2000.03918.x},
  319, 939

\bibitem[\protect\citeauthoryear{{Cooray} \& {Sheth}}{{Cooray} \&
  {Sheth}}{2002}]{Cooray:2002aa}
{Cooray} A.,  {Sheth} R.,  2002, \mn@doi [\physrep]
  {10.1016/S0370-1573(02)00276-4}, \href
  {http://adsabs.harvard.edu/abs/2002PhR...372....1C} {372, 1}

\bibitem[\protect\citeauthoryear{{Croft}, {Dalton}  \& {Efstathiou}}{{Croft}
  et~al.}{1999}]{Croft:1999aa}
{Croft} R.~A.~C.,  {Dalton} G.~B.,   {Efstathiou} G.,  1999, \mn@doi [\mnras]
  {10.1046/j.1365-8711.1999.02381.x}, \href
  {http://adsabs.harvard.edu/abs/1999MNRAS.305..547C} {305, 547}

\bibitem[\protect\citeauthoryear{{Das} et~al.,}{{Das}
  et~al.}{2014}]{Das:2014aa}
{Das} S.,  et~al., 2014, \mn@doi [\jcap] {10.1088/1475-7516/2014/04/014}, \href
  {http://adsabs.harvard.edu/abs/2014JCAP...04..014D} {4, 014}

\bibitem[\protect\citeauthoryear{{Di Dio}, {Montanari}, {Lesgourgues}  \&
  {Durrer}}{{Di Dio} et~al.}{2013}]{Di-Dio:2013aa}
{Di Dio} E.,  {Montanari} F.,  {Lesgourgues} J.,   {Durrer} R.,  2013, \mn@doi
  [\jcap] {10.1088/1475-7516/2013/11/044}, \href
  {http://adsabs.harvard.edu/abs/2013JCAP...11..044D} {11, 044}

\bibitem[\protect\citeauthoryear{{Efstathiou}}{{Efstathiou}}{2004}]{Efstathiou:2004aa}
{Efstathiou} G.,  2004, \mn@doi [\mnras] {10.1111/j.1365-2966.2004.07530.x},
  \href {http://adsabs.harvard.edu/abs/2004MNRAS.349..603E} {349, 603}

\bibitem[\protect\citeauthoryear{Eisenstein et~al.}{Eisenstein
  et~al.}{2011}]{Eisenstein:2011sa}
Eisenstein D.~J.,  et~al., 2011, \mn@doi [Astron.J.]
  {10.1088/0004-6256/142/3/72}, 142, 72

\bibitem[\protect\citeauthoryear{Estrada, Sefusatti  \& Frieman}{Estrada
  et~al.}{2009}]{Estrada:2008em}
Estrada J.,  Sefusatti E.,   Frieman J.~A.,  2009, \mn@doi [Astrophys.J.]
  {10.1088/0004-637X/692/1/265}, 692, 265

\bibitem[\protect\citeauthoryear{{Farahi}, {Evrard}, {Rozo}, {Rykoff}  \&
  {Wechsler}}{{Farahi} et~al.}{2016}]{Farahi:2016aa}
{Farahi} A.,  {Evrard} A.~E.,  {Rozo} E.,  {Rykoff} E.~S.,   {Wechsler} R.~H.,
  2016, preprint, \href {http://adsabs.harvard.edu/abs/2016arXiv160105773F} {}
  (\mn@eprint {arXiv} {1601.05773})

\bibitem[\protect\citeauthoryear{{Fedeli}, {Carbone}, {Moscardini}  \&
  {Cimatti}}{{Fedeli} et~al.}{2011}]{2011MNRAS.414.1545F}
{Fedeli} C.,  {Carbone} C.,  {Moscardini} L.,   {Cimatti} A.,  2011, \mn@doi
  [\mnras] {10.1111/j.1365-2966.2011.18490.x}, \href
  {http://adsabs.harvard.edu/abs/2011MNRAS.414.1545F} {414, 1545}

\bibitem[\protect\citeauthoryear{{Fry} \& {Gaztanaga}}{{Fry} \&
  {Gaztanaga}}{1993}]{Fry:1993aa}
{Fry} J.~N.,  {Gaztanaga} E.,  1993, \mn@doi [\apj] {10.1086/173015}, \href
  {http://adsabs.harvard.edu/abs/1993ApJ...413..447F} {413, 447}

\bibitem[\protect\citeauthoryear{{Gao} \& {White}}{{Gao} \&
  {White}}{2007}]{2007MNRAS.377L...5G}
{Gao} L.,  {White} S.~D.~M.,  2007, \mn@doi [\mnras]
  {10.1111/j.1745-3933.2007.00292.x}, \href
  {http://adsabs.harvard.edu/abs/2007MNRAS.377L...5G} {377, L5}

\bibitem[\protect\citeauthoryear{{Gao}, {Springel}  \& {White}}{{Gao}
  et~al.}{2005}]{2005MNRAS.363L..66G}
{Gao} L.,  {Springel} V.,   {White} S.~D.~M.,  2005, \mn@doi [\mnras]
  {10.1111/j.1745-3933.2005.00084.x}, \href
  {http://adsabs.harvard.edu/abs/2005MNRAS.363L..66G} {363, L66}

\bibitem[\protect\citeauthoryear{{Giannantonio} \& {Percival}}{{Giannantonio}
  \& {Percival}}{2014}]{Giannantonio2014}
{Giannantonio} T.,  {Percival} W.~J.,  2014, \mn@doi [\mnras]
  {10.1093/mnrasl/slu036}, \href
  {http://adsabs.harvard.edu/abs/2014MNRAS.441L..16G} {441, L16}

\bibitem[\protect\citeauthoryear{{Giannantonio} et~al.,}{{Giannantonio}
  et~al.}{2006}]{Giannantonio:2006aa}
{Giannantonio} T.,  et~al., 2006, \mn@doi [\prd] {10.1103/PhysRevD.74.063520},
  \href {http://adsabs.harvard.edu/abs/2006PhRvD..74f3520G} {74, 063520}

\bibitem[\protect\citeauthoryear{Giannantonio, Crittenden, Nichol  \&
  Ross}{Giannantonio et~al.}{2012}]{Giannantonio:2012aa}
Giannantonio T.,  Crittenden R.,  Nichol R.,   Ross A.~J.,  2012, \mn@doi
  [Mon.Not.Roy.Astron.Soc.] {10.1111/j.1365-2966.2012.21896.x}, 426, 2581

\bibitem[\protect\citeauthoryear{{Giannantonio}, {Fosalba}, {Cawthon}, {Omori},
  {Crocce}  et~al.}{{Giannantonio} et~al.}{2016}]{Giannantonio:2016aa}
{Giannantonio} T.,  {Fosalba} P.,  {Cawthon} R.,  {Omori} Y.,  {Crocce} M.,
  et~al., 2016, \mn@doi [\mnras] {10.1093/mnras/stv2678}, \href
  {http://adsabs.harvard.edu/abs/2016MNRAS.456.3213G} {456, 3213}

\bibitem[\protect\citeauthoryear{Gorski, Hivon, Banday, Wandelt, Hansen
  et~al.}{Gorski et~al.}{2005}]{Gorski:2004by}
Gorski K.,  Hivon E.,  Banday A.,  Wandelt B.,  Hansen F.,   et~al., 2005,
  \mn@doi [Astrophys.J.] {10.1086/427976}, 622, 759

\bibitem[\protect\citeauthoryear{{Grieb}, {S{\'a}nchez}, {Salazar-Albornoz},
  {Scoccimarro}, {Crocce}, {Dalla Vecchia}  \& et al.}{{Grieb}
  et~al.}{2016}]{Grieb:2016aa}
{Grieb} J.~N.,  {S{\'a}nchez} A.~G.,  {Salazar-Albornoz} S.,  {Scoccimarro} R.,
   {Crocce} M.,  {Dalla Vecchia} C.,   et al. 2016, preprint, \href
  {http://adsabs.harvard.edu/abs/2016arXiv160703143G} {} (\mn@eprint {arXiv}
  {1607.03143})

\bibitem[\protect\citeauthoryear{{Hamaus}, {Seljak}, {Desjacques}, {Smith}  \&
  {Baldauf}}{{Hamaus} et~al.}{2010}]{Hamaus:2010aa}
{Hamaus} N.,  {Seljak} U.,  {Desjacques} V.,  {Smith} R.~E.,   {Baldauf} T.,
  2010, \mn@doi [\prd] {10.1103/PhysRevD.82.043515}, \href
  {http://adsabs.harvard.edu/abs/2010PhRvD..82d3515H} {82, 043515}

\bibitem[\protect\citeauthoryear{{Hamaus}, {Wandelt}, {Sutter}, {Lavaux}  \&
  {Warren}}{{Hamaus} et~al.}{2014}]{Hamaus:2014aa}
{Hamaus} N.,  {Wandelt} B.~D.,  {Sutter} P.~M.,  {Lavaux} G.,   {Warren} M.~S.,
   2014, \mn@doi [Physical Review Letters] {10.1103/PhysRevLett.112.041304},
  \href {http://adsabs.harvard.edu/abs/2014PhRvL.112d1304H} {112, 041304}

\bibitem[\protect\citeauthoryear{{Hamaus}, {Pisani}, {Sutter}, {Lavaux},
  {Escoffier}, {Wandelt}  \& {Weller}}{{Hamaus} et~al.}{2016}]{Hamaus:2016aa}
{Hamaus} N.,  {Pisani} A.,  {Sutter} P.~M.,  {Lavaux} G.,  {Escoffier} S.,
  {Wandelt} B.~D.,   {Weller} J.,  2016, \mn@doi [Physical Review Letters]
  {10.1103/PhysRevLett.117.091302}, \href
  {http://adsabs.harvard.edu/abs/2016PhRvL.117i1302H} {117, 091302}

\bibitem[\protect\citeauthoryear{{Hartlap}, {Simon}  \& {Schneider}}{{Hartlap}
  et~al.}{2007}]{Hartlap:2007aa}
{Hartlap} J.,  {Simon} P.,   {Schneider} P.,  2007, \mn@doi [\aap]
  {10.1051/0004-6361:20066170}, \href
  {http://adsabs.harvard.edu/abs/2007A%26A...464..399H} {464, 399}

\bibitem[\protect\citeauthoryear{{Hayes}, {Brunner}  \& {Ross}}{{Hayes}
  et~al.}{2011}]{Hayes:2011aa}
{Hayes} B.,  {Brunner} R.,   {Ross} A.,  2011, preprint, \href
  {http://adsabs.harvard.edu/abs/2011arXiv1112.5723H} {} (\mn@eprint {arXiv}
  {1112.5723})

\bibitem[\protect\citeauthoryear{{Ho}, {Cuesta}, {Seo}  et~al.}{{Ho}
  et~al.}{2012}]{Ho:2012aa}
{Ho} S.,  {Cuesta} A.,  {Seo} H.-J.,   et~al., 2012, \mn@doi [\apj]
  {10.1088/0004-637X/761/1/14}, \href
  {http://adsabs.harvard.edu/abs/2012ApJ...761...14H} {761, 14}

\bibitem[\protect\citeauthoryear{Huetsi}{Huetsi}{2009}]{Huetsi:2009zq}
Huetsi G.,  2009

\bibitem[\protect\citeauthoryear{{H{\"u}tsi} \& {Lahav}}{{H{\"u}tsi} \&
  {Lahav}}{2008}]{Hutsi:2008aa}
{H{\"u}tsi} G.,  {Lahav} O.,  2008, \mn@doi [\aap]
  {10.1051/0004-6361:200810250}, \href
  {http://adsabs.harvard.edu/abs/2008A%26A...492..355H} {492, 355}

\bibitem[\protect\citeauthoryear{{Johnson}, {Blake}, {Dossett}, {Koda},
  {Parkinson}  \& {Joudaki}}{{Johnson} et~al.}{2016}]{Johnson:2016aa}
{Johnson} A.,  {Blake} C.,  {Dossett} J.,  {Koda} J.,  {Parkinson} D.,
  {Joudaki} S.,  2016, \mn@doi [\mnras] {10.1093/mnras/stw447}, \href
  {http://adsabs.harvard.edu/abs/2016MNRAS.458.2725J} {458, 2725}

\bibitem[\protect\citeauthoryear{{La Porta}, {Burigana}, {Reich}  \&
  {Reich}}{{La Porta} et~al.}{2008}]{La-Porta:2008aa}
{La Porta} L.,  {Burigana} C.,  {Reich} W.,   {Reich} P.,  2008, \mn@doi [\aap]
  {10.1051/0004-6361:20078435}, \href
  {http://adsabs.harvard.edu/abs/2008A%26A...479..641L} {479, 641}

\bibitem[\protect\citeauthoryear{{Lilje} \& {Efstathiou}}{{Lilje} \&
  {Efstathiou}}{1988}]{LiljeEfstathiou1988MNRAS}
{Lilje} P.~B.,  {Efstathiou} G.,  1988, Mon.Not.Roy.Astron.Soc., \href
  {http://adsabs.harvard.edu/abs/1988MNRAS.231..635L} {231, 635}

\bibitem[\protect\citeauthoryear{Mana, Giannantonio, Weller, Hoyle, Huetsi
  et~al.}{Mana et~al.}{2013}]{Mana:2013qba}
Mana A.,  Giannantonio T.,  Weller J.,  Hoyle B.,  Huetsi G.,   et~al., 2013,
  \mn@doi [Mon.Not.Roy.Astron.Soc.] {10.1093/mnras/stt1062}, 434, 684

\bibitem[\protect\citeauthoryear{{McDonald}}{{McDonald}}{2006}]{McDonald:2006aa}
{McDonald} P.,  2006, \mn@doi [\prd] {10.1103/PhysRevD.74.103512}, \href
  {http://adsabs.harvard.edu/abs/2006PhRvD..74j3512M} {74, 103512}

\bibitem[\protect\citeauthoryear{{Miyatake}, {More}, {Takada}, {Spergel},
  {Mandelbaum}, {Rykoff}  \& {Rozo}}{{Miyatake} et~al.}{2016}]{Miyatake:2016aa}
{Miyatake} H.,  {More} S.,  {Takada} M.,  {Spergel} D.~N.,  {Mandelbaum} R.,
  {Rykoff} E.~S.,   {Rozo} E.,  2016, \mn@doi [Physical Review Letters]
  {10.1103/PhysRevLett.116.041301}, \href
  {http://adsabs.harvard.edu/abs/2016PhRvL.116d1301M} {116, 041301}

\bibitem[\protect\citeauthoryear{{More} et~al.,}{{More}
  et~al.}{2016}]{More:2016aa}
{More} S.,  et~al., 2016, \mn@doi [\apj] {10.3847/0004-637X/825/1/39}, \href
  {http://adsabs.harvard.edu/abs/2016ApJ...825...39M} {825, 39}

\bibitem[\protect\citeauthoryear{{Padmanabhan}, {Xu}, {Eisenstein}, {Scalzo},
  {Cuesta}, {Mehta}  \& {Kazin}}{{Padmanabhan}
  et~al.}{2012}]{Padmanabhan:2012aa}
{Padmanabhan} N.,  {Xu} X.,  {Eisenstein} D.~J.,  {Scalzo} R.,  {Cuesta} A.~J.,
   {Mehta} K.~T.,   {Kazin} E.,  2012, \mn@doi [\mnras]
  {10.1111/j.1365-2966.2012.21888.x}, \href
  {http://adsabs.harvard.edu/abs/2012MNRAS.427.2132P} {427, 2132}

\bibitem[\protect\citeauthoryear{{Peebles}}{{Peebles}}{1974}]{Peebles1974ApJS}
{Peebles} P.~J.~E.,  1974, \mn@doi [Astrophys.J.Suppl.] {10.1086/190309}, \href
  {http://adsabs.harvard.edu/abs/1974ApJS...28...37P} {28, 37}

\bibitem[\protect\citeauthoryear{Percival et~al.}{Percival
  et~al.}{2001}]{Percival:2001hw}
Percival W.~J.,  et~al., 2001, \mn@doi [Mon.Not.Roy.Astron.Soc.]
  {10.1046/j.1365-8711.2001.04827.x}, 327, 1297

\bibitem[\protect\citeauthoryear{{Percival} et~al.,}{{Percival}
  et~al.}{2010}]{Percival:2010aa}
{Percival} W.~J.,  et~al., 2010, \mn@doi [\mnras]
  {10.1111/j.1365-2966.2009.15812.x}, \href
  {http://adsabs.harvard.edu/abs/2010MNRAS.401.2148P} {401, 2148}

\bibitem[\protect\citeauthoryear{{Pillepich}, {Porciani}  \&
  {Reiprich}}{{Pillepich} et~al.}{2012}]{2012MNRAS.422...44P}
{Pillepich} A.,  {Porciani} C.,   {Reiprich} T.~H.,  2012, \mn@doi [\mnras]
  {10.1111/j.1365-2966.2012.20443.x}, \href
  {http://adsabs.harvard.edu/abs/2012MNRAS.422...44P} {422, 44}

\bibitem[\protect\citeauthoryear{{Planck Collaboration} et~al.,}{{Planck
  Collaboration} et~al.}{2014}]{Planck-Collaboration:2014aa}
{Planck Collaboration} et~al., 2014, \mn@doi [\aap]
  {10.1051/0004-6361/201321591}, \href
  {http://adsabs.harvard.edu/abs/2014A%26A...571A..16P} {571, A16}

\bibitem[\protect\citeauthoryear{{Reichardt} et~al.,}{{Reichardt}
  et~al.}{2012}]{Reichardt:2012aa}
{Reichardt} C.~L.,  et~al., 2012, \mn@doi [\apj] {10.1088/0004-637X/755/1/70},
  \href {http://adsabs.harvard.edu/abs/2012ApJ...755...70R} {755, 70}

\bibitem[\protect\citeauthoryear{{Ross}, {Ho}, {Cuesta}, {Tojeiro}, {Percival},
  {Wake}  \& {Masters}}{{Ross} et~al.}{2011}]{Ross:2011aa}
{Ross} A.~J.,  {Ho} S.,  {Cuesta} A.~J.,  {Tojeiro} R.,  {Percival} W.~J.,
  {Wake} D.,   {Masters} 2011, \mn@doi [\mnras]
  {10.1111/j.1365-2966.2011.19351.x}, \href
  {http://adsabs.harvard.edu/abs/2011MNRAS.417.1350R} {417, 1350}

\bibitem[\protect\citeauthoryear{Rykoff et~al.}{Rykoff
  et~al.}{2014}]{Rykoff:2013ovv}
Rykoff E.,  et~al., 2014, \mn@doi [Astrophys.J.] {10.1088/0004-637X/785/2/104},
  785, 104

\bibitem[\protect\citeauthoryear{{S{\'a}nchez}, {Lambas}, {B{\"o}hringer}  \&
  {Schuecker}}{{S{\'a}nchez} et~al.}{2005}]{2005MNRAS.362.1225S}
{S{\'a}nchez} A.~G.,  {Lambas} D.~G.,  {B{\"o}hringer} H.,   {Schuecker} P.,
  2005, \mn@doi [\mnras] {10.1111/j.1365-2966.2005.09377.x}, \href
  {http://adsabs.harvard.edu/abs/2005MNRAS.362.1225S} {362, 1225}

\bibitem[\protect\citeauthoryear{{Schlegel}, {Finkbeiner}  \&
  {Davis}}{{Schlegel} et~al.}{1998}]{Schlegel:1998lr}
{Schlegel} D.~J.,  {Finkbeiner} D.~P.,   {Davis} M.,  1998, \mn@doi [\apj]
  {10.1086/305772}, \href {http://adsabs.harvard.edu/abs/1998ApJ...500..525S}
  {500, 525}

\bibitem[\protect\citeauthoryear{{Schuecker}, {B{\"o}hringer}, {Collins}  \&
  {Guzzo}}{{Schuecker} et~al.}{2003}]{Schuecker:2003aa}
{Schuecker} P.,  {B{\"o}hringer} H.,  {Collins} C.~A.,   {Guzzo} L.,  2003,
  \mn@doi [\aap] {10.1051/0004-6361:20021715}, \href
  {http://adsabs.harvard.edu/abs/2003A%26A...398..867S} {398, 867}

\bibitem[\protect\citeauthoryear{{Seldner} \& {Peebles}}{{Seldner} \&
  {Peebles}}{1977a}]{SeldnerPeebles1977Apjla}
{Seldner} M.,  {Peebles} P.~J.~E.,  1977a, \mn@doi [Astrophys.J.Letters]
  {10.1086/182428}, \href {http://adsabs.harvard.edu/abs/1977ApJ...214L...1S}
  {214, L1}

\bibitem[\protect\citeauthoryear{{Seldner} \& {Peebles}}{{Seldner} \&
  {Peebles}}{1977b}]{SeldnerPeebles1977ApJb}
{Seldner} M.,  {Peebles} P.~J.~E.,  1977b, \mn@doi [Astrophys.J.]
  {10.1086/155404}, \href {http://adsabs.harvard.edu/abs/1977ApJ...215..703S}
  {215, 703}

\bibitem[\protect\citeauthoryear{{Seldner}, {Siebers}, {Groth}  \&
  {Peebles}}{{Seldner} et~al.}{1977}]{1977AJ.....82..249S}
{Seldner} M.,  {Siebers} B.,  {Groth} E.~J.,   {Peebles} P.~J.~E.,  1977,
  \mn@doi [Astronomical Journal] {10.1086/112039}, \href
  {http://adsabs.harvard.edu/abs/1977AJ.....82..249S} {82, 249}

\bibitem[\protect\citeauthoryear{{Seljak}}{{Seljak}}{2000}]{Seljak:2000aa}
{Seljak} U.,  2000, \mn@doi [\mnras] {10.1046/j.1365-8711.2000.03715.x}, \href
  {http://adsabs.harvard.edu/abs/2000MNRAS.318..203S} {318, 203}

\bibitem[\protect\citeauthoryear{{Shane} \& {Wirtanen}}{{Shane} \&
  {Wirtanen}}{1967}]{Shane1967}
{Shane} C.~D.,  {Wirtanen} C.~A.,  1967, Publications of the Lick Observatory,
  22, part 1

\bibitem[\protect\citeauthoryear{{Sheth} \& {Tormen}}{{Sheth} \&
  {Tormen}}{2004}]{2004MNRAS.350.1385S}
{Sheth} R.~K.,  {Tormen} G.,  2004, \mn@doi [\mnras]
  {10.1111/j.1365-2966.2004.07733.x}, \href
  {http://adsabs.harvard.edu/abs/2004MNRAS.350.1385S} {350, 1385}

\bibitem[\protect\citeauthoryear{{Simet}, {McClintock}, {Mandelbaum}, {Rozo},
  {Rykoff}, {Sheldon}  \& {Wechsler}}{{Simet} et~al.}{2016}]{Simet:2016aa}
{Simet} M.,  {McClintock} T.,  {Mandelbaum} R.,  {Rozo} E.,  {Rykoff} E.,
  {Sheldon} E.,   {Wechsler} R.~H.,  2016, preprint, \href
  {http://adsabs.harvard.edu/abs/2016arXiv160306953S} {} (\mn@eprint {arXiv}
  {1603.06953})

\bibitem[\protect\citeauthoryear{{Smith} et~al.,}{{Smith}
  et~al.}{2003}]{Smith:2003aa}
{Smith} R.~E.,  et~al., 2003, \mn@doi [\mnras]
  {10.1046/j.1365-8711.2003.06503.x}, \href
  {http://adsabs.harvard.edu/abs/2003MNRAS.341.1311S} {341, 1311}

\bibitem[\protect\citeauthoryear{Szapudi, Prunet, Pogosyan, Szalay  \&
  Bond}{Szapudi et~al.}{2000}]{Szapudi:2000xj}
Szapudi I.,  Prunet S.,  Pogosyan D.,  Szalay A.~S.,   Bond J.~R.,  2000

\bibitem[\protect\citeauthoryear{{Taylor}, {Joachimi}  \& {Kitching}}{{Taylor}
  et~al.}{2013}]{Taylor:2013aa}
{Taylor} A.,  {Joachimi} B.,   {Kitching} T.,  2013, \mn@doi [\mnras]
  {10.1093/mnras/stt270}, \href
  {http://adsabs.harvard.edu/abs/2013MNRAS.432.1928T} {432, 1928}

\bibitem[\protect\citeauthoryear{Tegmark et~al.}{Tegmark
  et~al.}{2004}]{Tegmark:2003ud}
Tegmark M.,  et~al., 2004, \mn@doi [Phys.Rev.] {10.1103/PhysRevD.69.103501},
  D69, 103501

\bibitem[\protect\citeauthoryear{{Tinker}, {Kravtsov}, {Klypin}, {Abazajian},
  {Warren}, {Yepes}, {Gottl{\"o}ber}  \& {Holz}}{{Tinker}
  et~al.}{2008}]{Tinker:2008aa}
{Tinker} J.,  {Kravtsov} A.~V.,  {Klypin} A.,  {Abazajian} K.,  {Warren} M.,
  {Yepes} G.,  {Gottl{\"o}ber} S.,   {Holz} D.~E.,  2008, \mn@doi [\apj]
  {10.1086/591439}, \href {http://adsabs.harvard.edu/abs/2008ApJ...688..709T}
  {688, 709}

\bibitem[\protect\citeauthoryear{{Tinker}, {Robertson}, {Kravtsov}, {Klypin},
  {Warren}, {Yepes}  \& {Gottl{\"o}ber}}{{Tinker} et~al.}{2010}]{Tinker:2010aa}
{Tinker} J.~L.,  {Robertson} B.~E.,  {Kravtsov} A.~V.,  {Klypin} A.,  {Warren}
  M.~S.,  {Yepes} G.,   {Gottl{\"o}ber} S.,  2010, \mn@doi [\apj]
  {10.1088/0004-637X/724/2/878}, \href
  {http://adsabs.harvard.edu/abs/2010ApJ...724..878T} {724, 878}

\bibitem[\protect\citeauthoryear{{Veropalumbo}, {Marulli}, {Moscardini},
  {Moresco}  \& {Cimatti}}{{Veropalumbo} et~al.}{2016}]{Veropalumbo:2016aa}
{Veropalumbo} A.,  {Marulli} F.,  {Moscardini} L.,  {Moresco} M.,   {Cimatti}
  A.,  2016, \mn@doi [\mnras] {10.1093/mnras/stw306}, \href
  {http://adsabs.harvard.edu/abs/2016MNRAS.458.1909V} {458, 1909}

\bibitem[\protect\citeauthoryear{{Wechsler}, {Zentner}, {Bullock}, {Kravtsov}
  \& {Allgood}}{{Wechsler} et~al.}{2006}]{2006ApJ...652...71W}
{Wechsler} R.~H.,  {Zentner} A.~R.,  {Bullock} J.~S.,  {Kravtsov} A.~V.,
  {Allgood} B.,  2006, \mn@doi [\apj] {10.1086/507120}, \href
  {http://adsabs.harvard.edu/abs/2006ApJ...652...71W} {652, 71}

\bibitem[\protect\citeauthoryear{{York} et~al.,}{{York}
  et~al.}{2000}]{York2000}
{York} D.~G.,  et~al., 2000, \mn@doi [\aj] {10.1086/301513}, \href
  {http://adsabs.harvard.edu/abs/2000AJ....120.1579Y} {120, 1579}

\bibitem[\protect\citeauthoryear{{Zhao}, {Saito}, {Percival}  et~al.}{{Zhao}
  et~al.}{2013}]{Zhao:2013aa}
{Zhao} G.-B.,  {Saito} S.,  {Percival} W.~J.,   et~al., 2013, \mn@doi [\mnras]
  {10.1093/mnras/stt1710}, \href
  {http://adsabs.harvard.edu/abs/2013MNRAS.436.2038Z} {436, 2038}

\bibitem[\protect\citeauthoryear{{Zu} \& {Weinberg}}{{Zu} \&
  {Weinberg}}{2013}]{2013MNRAS.431.3319Z}
{Zu} Y.,  {Weinberg} D.~H.,  2013, \mn@doi [\mnras] {10.1093/mnras/stt411},
  \href {http://adsabs.harvard.edu/abs/2013MNRAS.431.3319Z} {431, 3319}

\makeatother
\end{thebibliography}

\label{lastpage}

\end{document}